%% file: main.tex
\documentclass{article}
\usepackage{arxiv}

\usepackage[utf8]{inputenc} % allow utf-8 input
\usepackage[T1]{fontenc}    % use 8-bit T1 fonts
\usepackage{hyperref}       % hyperlinks
\usepackage{url}            % simple URL typesetting
\usepackage{booktabs}       % professional-quality tables
\usepackage{amsfonts}       % blackboard math symbols
\usepackage{nicefrac}       % compact symbols for 1/2, etc.
\usepackage{microtype}      % microtypography
\usepackage{lipsum}		% Can be removed after putting your text content
\usepackage{graphicx}
\usepackage{doi}

\usepackage{textcomp}

\usepackage{tikz}
\usepackage{booktabs}
\usepackage{optidef}
\usepackage{bm}

\usepackage{graphicx}
\usepackage{amsmath}
\usepackage[version=4]{mhchem}
\usepackage{siunitx}
\usepackage{longtable,tabularx}
\title{Multi-Target Spacecraft Mission Design using Convex Optimization and Binary Integer Programming}

\date{}

\usepackage{authblk}

\author[1]{%
	Jack Yarndley\thanks{\texttt{jyar540@aucklanduni.ac.nz}}\hspace{1.2mm}%
}
\author[1]{%
	{Harry Holt}%
}
\author[1]{%
	{Roberto Armellin}%
}

\affil[1]{Te P\=unaha \=Atea -- Space Institute, University of Auckland, Auckland 1010, New Zealand}

\renewcommand{\headeright}{}
\renewcommand{\undertitle}{Preprint}
\renewcommand{\shorttitle}{\textit{Multi-Target Spacecraft Mission Design using Convex Optimization and Binary Integer Programming}}

\hypersetup{
pdftitle={Multi-Target Spacecraft Mission Design using Convex Optimization and Binary Integer Programming},
pdfsubject={physics.space-ph},
pdfauthor={Jack Yarndley, Harry Holt, Roberto Armellin},
pdfkeywords={trajectory optimization, space, asteroid mining, convex programming, binary programming},
}

\usepackage[numbers]{natbib}

\begin{document}

\maketitle

\begin{abstract}
The optimal design of multi-target rendezvous and flyby missions is challenging due to the combination of traditional spacecraft trajectory optimization and high-dimensional combinatorial problems. This often requires large-scale global search techniques or simplified approximations that rely on manual tuning to be performant. While global search techniques are typically computationally expensive, limiting their use in time- or cost-constrained scenarios, this work proposes a computationally efficient nested-loop approach. The problem is split into separate combinatorial and optimal control subproblems: the combinatorial problem is solved using Binary Integer Programming (BIP) with a fixed rendezvous time schedule, while the optimal control problem is handled with adaptive-mesh Sequential Convex Programming (SCP), which also optimizes the time schedule. By iterating these processes in a nested-loop structure, the approach can efficiently find high-quality solutions. When, applied to the Global Trajectory Optimization Competition 12 (GTOC 12) problem, this method results in several new best-known solutions.
\end{abstract}
\section{\label{introduction}Introduction}
Multi-target mission design represents a growing trend within the fields of space exploration and exploitation. Traditionally, spacecraft trajectory design has focused on missions with a single primary target or objective \cite{hallPioneer10111974, reinhardGiottoEncounterComet1986, schmidtMarsExpressESA2003a}. However, recent missions, such as the recently launched JUICE \cite{grassetJUpiterICyMoons2013} and Lucy \cite{englanderTrajectoryDesignLucy2019} missions, have begun to adopt the concept of targeting multiple primary objectives. In the literature, this class of problems is often exemplified by asteroid tour design \cite{izzoAutomatedAsteroidSelection2007, longStochasticMissionExploration2024, dicarloLowthrustTourMain2018}, where a single spacecraft performs rendezvous or flybys with multiple asteroids along its trajectory. The design of such multi-target missions is important because they typically offer significant reductions in the cost to visit each target when compared to sequences of single-target missions. Consequently, these problems have been frequently proposed in spacecraft optimization challenges, such as the Global Trajectory Optimization Competition (GTOC) \cite{izzo1stACTGlobal2007}.

The design requirements of multi-target missions introduce a considerable increase in complexity over single-target missions. This is primarily because, in addition to optimizing the trajectory, the selection and sequencing of the targets must be integrated into the optimization process. Such problems are frequently encoded as Mixed-Integer Non-Linear Programming (MINLP) \cite{schlueterNonlinearMixedInteger2012} problems, which are also known as Hybrid Optimal Control Programming (HOCP) \cite{rossHybridOptimalControl2005} problems. In these formulations, a set of integer variables is used to define the sequence of targets to visit, whilst continuous variables handle the optimization of the control parameters. Although significant efforts within the literature have been put into directly solving problems in this encoding, their solutions are often difficult to obtain and are computationally expensive when compared to alternative solution methodologies \cite{dambrosioMixedIntegerNonlinear2013}.

A common strategy to mitigate the difficulties of solving MINLP problems is to partition the combinatorial ordering problem (the integer variables) from the optimal control problem (the continuous variables) and solve them in a sequential manner. However, this approach introduces challenges due to the strong coupling often encountered between the two problems. For instance, in asteroid tour design, the optimal sequence of targets directly influences the optimal control profile and vice versa. One method that attempts to address this issue is to employ a nested optimization process \cite{englanderAutomatedMissionPlanning2012}, which iteratively alternates between solving the combinatorial and optimal control problems, rather than handling them in strict sequence. Thus, developing efficient solution methods for both the combinatorial and optimal control problems is key to enabling an effective nested optimization process.

The combinatorial ordering problem closely resembles the well-known Travelling Salesman Problem (TSP) \cite{lawlerTravelingSalesmanProblem1985}, although with additional complexity. Consequently, a graph-based transcription is commonly employed, which allows for the use of a variety of search techniques that make use of this transcription. Depth-first and breadth-first search methods \cite{cormenIntroductionAlgorithmsFourth2022} systematically enumerate the entire search space, ensuring that once completed, they will identify the optimal combination along with all feasible solutions. However, the complete enumeration of the search space is often intractable, which can be partially addressed by branch and bound \cite{lawlerBranchandBoundMethodsSurvey1966} and its extensions, notably branch and cut \cite{padbergBranchCutAlgorithmResolution1991}. These methods utilize a similar tree structure to depth- and breadth-first search methods but prune the search space by eliminating branches when suboptimality or infeasibility can be established. Many leading commercial solvers, including Gurobi and CPLEX, implement advanced variations of this approach to tackle Mixed-Integer Programming (MIP) problems \cite{anandComparativeAnalysisOptimization2017}, of which Binary Integer Programming (BIP) is a subset.

As the size of the combinatorial problem grows, the additional complexity often makes it challenging (and sometimes intractable) to work with the complete problem. This leads to a clear preference towards incomplete search methods. Branch and bound (and branch and cut) methods can be employed with early termination, where a comparison between primal and dual objectives helps to quantify suboptimality relative to the global solution \cite{lawlerBranchandBoundMethodsSurvey1966}. Beam search is also widely used; it follows a breadth-first graph search strategy but limits computational effort by ranking branches at each level and expanding only a select number of the best options. This method is notable in that it has been successfully applied across many iterations of GTOC \cite{zhangGTOC11Results2023}. Dynamic programming can also be used, which is advantageous in that it directly leverages the optimal substructure property, allowing combinatorial problems to be split into smaller problems for which the optimal solution is the combination \cite{bellomeTrajectoryDesignMultiTarget2022}. For the most difficult combinatorial problems, stochastic algorithms are often preferred due to their flexibility and robustness across large classes of problems \cite{blumMetaheuristicsCombinatorialOptimization2003}. Stochastic algorithms include Differential Evolution (DE), Particle Swarm Optimization (PSO), Simulated Annealing (SA), Genetic Algorithms (GA), and Ant Colony Optimization (ACO), all of which have been applied with varying degrees of success to spacecraft tour design problems \cite{conwaySpacecraftTrajectoryOptimization2010}. 

The optimal control problem is also often challenging to solve and is generally considered to be a complex task. It is typically addressed through methodologies from the field of optimal control, which are well-established in the literature and have been applied across a wide range of engineering problems. Methods to solve optimal control problems can generally be classified into direct, indirect, and heuristic methods \cite{chaiReviewOptimizationTechniques2019}. 

Sequential Convex Programming (SCP) is a direct optimization method that has been widely applied to various aerospace applications, including launch vehicle ascent guidance \cite{benedikterConvexApproachThreeDimensional2021}, powered descent guidance \cite{kwonSequentialConvexProgramming2021}, and low-thrust multi-revolution interplanetary transfers \cite{hofmannComputationalGuidanceLowThrust2023a}. Numerous modifications to the core SCP algorithm have been proposed to improve performance, including those that explore different discretization strategies. Notably, adaptive-mesh SCP \cite{kumagaiAdaptiveMeshSequentialConvex2024} incorporates the time variable within the dynamical linearization and is therefore introduced into the convex approximation of the NLP. This adaptation allows for the mesh generated from the discretization (based on the direct transcription) to adjust dynamically, enabling superior optimality to be achieved with fewer transcription segments. However, for spacecraft optimal control problems, this approach presents a significant drawback: the introduction of the time variable can substantially degrade the quality of the dynamical linearization \cite{kumagaiAdaptiveMeshSequentialConvex2024}. As a result, the use of trust regions and additional iterations of SCP becomes necessary.

This paper introduces a nested-loop approach for solving multi-target spacecraft trajectory optimization problems using a combination of BIP and SCP. Firstly, the combinatorial aspect of the problem is solved using a BIP with a similar form to minimum-cost flow problems \cite{ahujaNetworkFlowsTheory1993}. An underlying graph structure is developed where the rendezvous of specific targets is represented by nodes and the intermediate transfers by arcs. The objective is to minimize transfer costs, which are computed in a discrete manner by assuming a fixed time schedule and approximating the cost via Lambert transfers. The BIP is constructed by introducing binary optimization variables for each arc, along with the necessary constraints, to ensure a single continuous trajectory is selected in the desired form. Following the solution to the combinatorial problem, the optimal control profile and rendezvous timing are simultaneously found using an adaptive-mesh SCP formulation, which has been substantially extended to permit many trajectories to be optimized at once. Once the optimal rendezvous timing has been found, the combinatorial aspect of the problem is resolved, and the process is iterated until convergence is achieved, completing the nested-loop approach.

The proposed approach forms a central component of our strategy to identify a range of high-quality post-competition solutions for the problem proposed in the 12th edition of GTOC \cite{gtoc12ProblemDescription}. This problem entails finding an optimal campaign of low-thrust ship trajectories, where each ship visits a sequence of asteroids to deploy miners and collect mined mass from the asteroids. The number of ships in a campaign is dependent on the total amount of mined mass, which, in turn, depends on the duration mining each asteroid. A trade-off must be struck between time, fuel consumption, and the number of asteroids visited. The winning solution by JPL presented a campaign of 35 ships that collectively mined mass from 313 asteroids, resulting in 626 rendezvous events. This clearly establishes the substantial scale necessary for high-quality solutions to the problem. 

\section{\label{problem_statement}Problem Statement}

\begin{table}
    \centering
    \caption{Problem specific parameters of GTOC 12.}
    \begin{tabular}{lc}\toprule
        \textbf{Parameter} & \textbf{Value} \\
        \midrule
        Minimum mission start time & 64328 MJD\\
        Maximum mission end time & 69807 MJD\\
        Maximum Earth starting velocity & 6 km/s\\
        Maximum Earth return velocity & 6 km/s\\
        Maximum ship mass (wet mass, $m_{\text{max}}$) & 3000 kg\\
        Minimum ship mass (dry mass, $m_{\text{min}}$) & 500 kg\\
        Miner mass ($m_{\text{miner}}$) & 40 kg\\
        Mining rate ($m_{\text{rate}}$) & 10 kg/yr\\
        Engine specific impulse ($I_{\text{sp}}$) & 4000 s\\
        Engine maximum thrust ($T_{\text{max}}$) & 0.6 N\\
        \bottomrule
    \end{tabular}
    \label{gtoc_parameters}
\end{table}

\begin{figure}
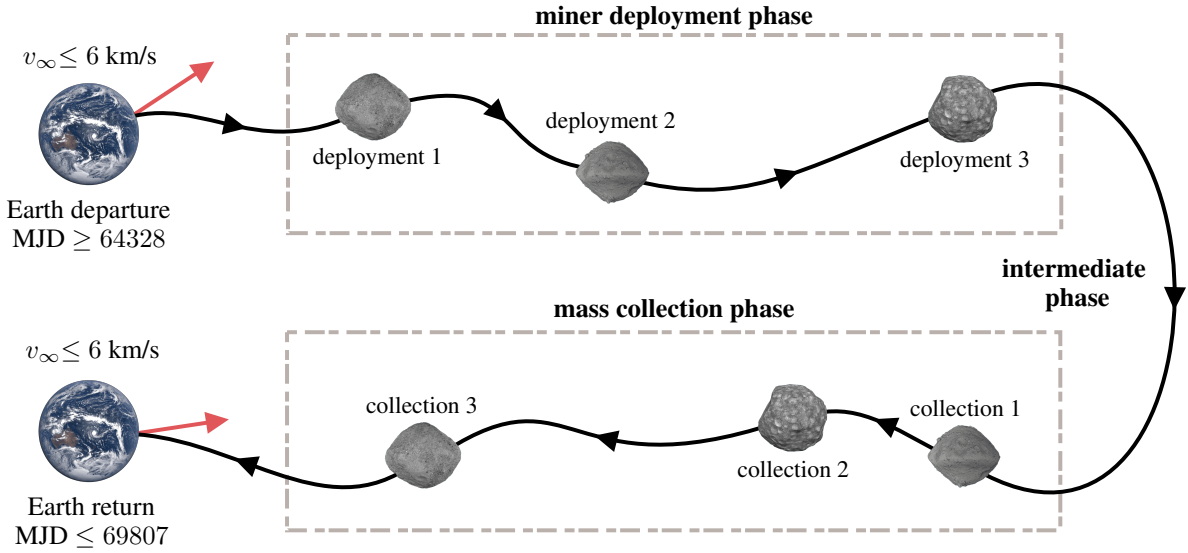

    \centering
    \include{figures/gtoc_explainer}
    \caption{Schematic of a valid ship trajectory for the problem proposed in GTOC 12.}
    \label{gtoc12_explainer}
\end{figure}

The 12th edition of GTOC (GTOC 12) \cite{gtoc12ProblemDescription} presented a challenging multi-target spacecraft trajectory design problem titled {\it sustainable asteroid mining}. This problem required designing a campaign of low-thrust spacecraft trajectories (ships) throughout the asteroid belt with the objective of mining as much mass as possible. A brief overview of the problem is provided below.

A catalog of 60,000 asteroids is provided as potential targets, resulting in an extremely large combinatorial problem size for potential trajectories. Important parameters related to the spacecraft and dynamics of the problem are detailed in Table~\ref{gtoc_parameters}. It is worth noting that the spacecraft engine parameters are somewhat unrealistic compared to current technology - particularly the 4000 $s$ $I_{\text{sp}}$ and corresponding thrust of 0.6 $N$. However, these parameters were designed to represent the future possibilities at the mission start time of 64328 MJD (January 1, 2035).

The objective of the problem is to maximize the total mass mined from the asteroids. Each ship must launch from Earth within a specified timeframe and can rendezvous with asteroids, where miners can be deployed. After some time, a ship can rendezvous with the same asteroid again to collect the mined mass before returning to Earth. A ship only contributes to the objective once it has returned to Earth with the collected mass. An example of the events throughout a ship trajectory is demonstrated in Figure~\ref{gtoc12_explainer}. In this example, a ship deploys miners and collects mined mass from three asteroids. It is important to note that the order of deployment and collection events does not need to follow the same sequence. 

Crucially, it is permitted for a coordinated campaign of ships to be launched, and the careful selection of ship trajectories that do not conflict becomes a problem. There is no requirement for deployment and collection events to be performed by the same ship. For instance, a miner can be deployed by one ship and the mined mass collected by another. The use of such strategies was a key factor that distinguished teams' solutions within the competition. Therefore, ships that both deploy and collect from the same asteroids are referred to as {\it self-cleaning} (similar to the trajectory seen in Figure~\ref{gtoc12_explainer}), and those that employ collaborative strategies are referred to as {\it mixed}. 

The number of ships allowed within a campaign depends on the average mined mass collected per ship, making ships that rendezvous faster and visit more asteroids highly advantageous. This behavior is described by the formula:
\begin{align} \label{number_ships}
    N \leq \min \left( 100, 2 e^{0.004 \bar{M}} \right)
\end{align}
where $N$ is the number of ships within a campaign and $\bar{M}$ is the average mined mass returned to Earth by each ship. Consequently, small increases in average mined mass per ship can significantly impact the overall campaign objective, by enabling the inclusion of whole additional ships into the campaign. 

The dynamics of the ships follow Keplerian motion with low-thrust control. With the state $\mathbf{x}=[\mathbf{r}, \mathbf{v}, m]$ and control $\mathbf{u}=[\mathbf{T}]$ the dynamical equations are as follows:
\begin{align} \label{dynamical_equation}
    \dot{\mathbf{x}} = f(\mathbf{x}, \mathbf{u}) = \left\{\begin{array}{l}
            \dot{\mathbf{r}} = \mathbf{v} \\
            \dot{\mathbf{v}} = -\frac{\mu}{r^3} \mathbf{r} + \frac{T_{\text{max}}}{m} \mathbf{T}\\
            \dot{m} = -\frac{T_{\text{max}}}{I_{\text{sp}} g_0} T
        \end{array}\right.
\end{align}
where $\mu$ is the gravitational constant of the Sun, $m$ the mining ship mass, and $0 \leq T \leq 1$ the thrust of the mining ship, normalized to $T_{\text{max}}$.

Additionally, the amount of mined mass is directly proportional to the amount of time that a miner has been deployed at the asteroid:
\begin{align} \label{mined_mass}
    m_{\text{mined}} = m_{\text{rate}} \left( t_{\text{collection}} - t_{\text{deployment}} \right)
\end{align}
where $t_{\text{deployment}}$ and $ t_{\text{collection}}$ represent the time of miner deployment and mass collection, respectively. As a result, an alternative expression for the objective of a campaign can be expressed as maximizing the total time between miner deployments and collections.

\section{\label{methodology}Methodology}

This section outlines the proposed approach for finding high-quality solutions to multi-target spacecraft trajectory design problems and then, separately, the assembly of many of these solutions into a coordinated campaign. As a preface, whilst many of the constraints and objectives are specifically tailored to the GTOC 12 problem, this approach is highly applicable beyond this context and can be easily generalized to a wide range of spacecraft trajectory design problems.

A central element to this approach is the use of a nested-loop structure, which is used to facilitate the transfer of updated information between the different sub-problems. By iteratively solving computationally tractable sub-problems rather than approaching the full problem at once, high-quality solutions can be obtained from the problem in an efficient manner. For this process, the sub-problems are constructed from the natural division of the full problem into what are labelled the combinatorial problem and the optimal control/rendezvous time problem. 

\begin{figure}
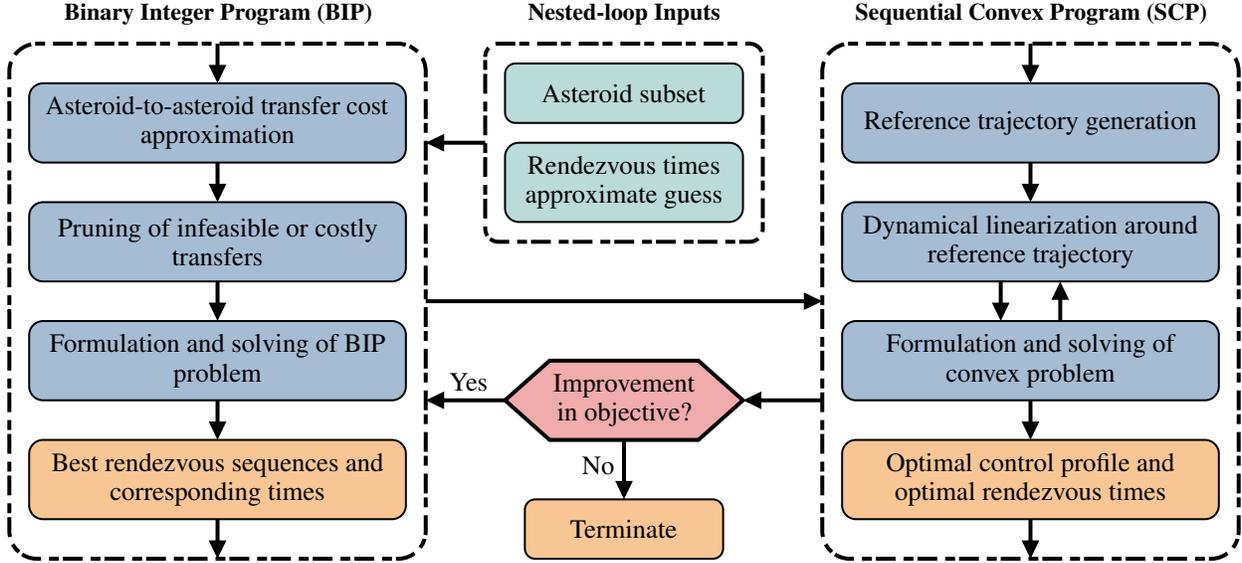

    \centering
    \include{figures/flowchart}
    \caption{Schematic of the nested-loop process between the BIP and SCP.}
    \label{nested_flow}
\end{figure}

Figure~\ref{nested_flow} outlines the overall structure of the solution process. The main inputs to the process are an asteroid \textit{subset} restricting the search to a limited number of asteroids out of the full data set and an initial estimate of the times for asteroid rendezvous to occur. These inputs are then transferred into a BIP to solve the combinatorial problem and determine the optimal sequence of asteroid rendezvous for each spacecraft. Subsequently, this sequence is transferred to the optimal control/rendezvous time problem, which is solved using SCP. The output of the SCP step is the optimal control profile and the corresponding optimal rendezvous times. Since the optimal rendezvous times are likely to differ from the initial estimates (or even those from previous iterations), they are passed back to the BIP, and the process is repeated to form the nested-loop structure, until no further improvements in the solution are found. 

Clearly, the assumptions that are made during the division of the full problem into sub-problems will significantly impact the optimality of the process. Therefore, many of these decisions have been made with the GTOC 12 problem in mind. The most impactful assumption is the restriction of the search space through the use of asteroid subsets. Given that the GTOC 12 problem presents 60,000 possible targets for rendezvous, the combinatorial problem easily becomes intractable in its full form. However, from this full set of possible targets, only a small number are good candidates for rendezvous. These tend to be well-connected and easily accessible from other asteroids. To isolate groups of these asteroids, a pruning and grouping process is applied, forming the {\it subsets}. This process evaluates the average Lambert transfer costs over the problem timeframe and is detailed in \cite{armellinGTOC12Results2024}. As a result, smaller groups of asteroids ranging in size from approximately 10 to 300 asteroids are created, making the combinatorial problem more tractable.

Additionally, an initial guess must be made for the rendezvous times. The selection of this initial guess is problem-specific and can be informed by the problem constraints. However, the initial guess does not need to be extremely accurate due to the optimization step in the SCP. For the GTOC 12 problem, our analysis determined that a rendezvous spacing of approximately 150-250 days would be appropriate for high-quality solutions \cite{armellinGTOC12Results2024}. This insight is used to derive an initial guess of the rendezvous times. Moreover, the start and end times of the rendezvous sequence are approximated by averaging the time of arrival and departure for optimal Lambert transfers from and to Earth for all asteroids within each subset.

Finally, an approximation is made to the asteroid-to-asteroid transfer costs by using the $\Delta v$ calculated via Lambert's problem, rather than the true mass cost. This approach offers two key advantages: solutions to Lambert's problem are very fast to calculate, and this estimate is independent of the spacecraft mass, which can only be approximated at each stage because it depends on the spacecraft's trajectory history. Because of this approximation, the output from the combinatorial problem may not perfectly match the true optimal ordering. Therefore, to improve the exploration of the algorithm, several of the best combinatorial solutions are refined in each stage of the nested-loop process.

\begin{table}
    \centering
    \caption{Classical orbital elements of the asteroids in the example subset at 64328 MJD.}

    \begin{tabular}{lrrrrrr}
    \toprule
    \textbf{ID} & $\mathbf{a} \left[\textbf{AU}\right]$ & $\mathbf{e} \left[\textbf{nd}\right]$ & $\mathbf{i} \left[\textbf{deg}\right]$ & $\mathbf{\Omega} \left[\textbf{deg}\right]$ & $\mathbf{\omega} \left[\textbf{deg}\right]$ & $\mathbf{M} \left[\textbf{deg}\right]$ \\
    \midrule
    $3241$ & $2.793$ & $0.038$ & $5.910$ & $81.870$ & $315.380$ & $253.512$ \\
    $15184$ & $2.777$ & $0.086$ & $1.620$ & $73.910$ & $295.490$ & $276.877$ \\
    $19702$ & $2.786$ & $0.069$ & $4.260$ & $184.720$ & $219.220$ & $244.407$ \\
    $46418$ & $2.777$ & $0.076$ & $4.760$ & $173.380$ & $225.120$ & $248.987$ \\
    $53592$ & $2.810$ & $0.060$ & $4.540$ & $175.630$ & $243.470$ & $237.719$ \\
    \bottomrule
    \end{tabular}
    \label{restricted_parameters}
\end{table}

\begin{table}
    \centering
    \caption{Initial guess for the transfer times within the example subset.}
    \begin{tabular}{lrr}\toprule
        \textbf{Event} & \textbf{Time} $\left[\textbf{MJD}\right]$ & $\mathbf{\Delta t} \left[\textbf{days}\right]$\\
        \midrule
        Earth Departure & $64438.00$ & - \\
        Miner Deployment 1 & $65038.00$ & $600.00$\\
        Miner Deployment 2 & $65213.00$ & $175.00$\\ 
        Miner Deployment 3 & $65388.00$ & $175.00$\\
        Mass Collection 1 & $68722.00$ & $3334.00$\\
        Mass Collection 2 & $68897.00$ & $175.00$ \\ 
        Mass Collection 3 & $69072.00$ & $175.00$\\
        Earth Arrival & $69772.00$ & $700.00$\\
        \bottomrule
    \end{tabular}
    \label{transfer_times_restricted}
\end{table}

As a demonstration example of the methodology, we introduce a restricted case of a GTOC 12 trajectory optimization problem. This exemplar case uses a highly restricted subset of 5 asteroids, which will enable a detailed analysis of the behavior of each stage of the process. The orbital elements of the asteroids in this subset, listed in Table~\ref{restricted_parameters}, are all clearly very similar. In the full problem, similar characteristics for the asteroids within a subset would be expected, as having similar orbital elements would permit the optimal transfer costs to be kept low and, subsequently, for the transfer times to be reduced.

The initial rendezvous time schedule used for the example problem is outlined in Figure~\ref{transfer_times_restricted}. A total of 3 asteroid deployment and collection events are designed, and in this case, if a feasible trajectory is found that meets the GTOC 12 constraints, it would yield a maximum of 302.59 kg of mined mass. This time schedule is calculated with a fixed rendezvous spacing of 175 days, and the initial and final transfer times are approximated based on optimal Lambert transfer durations. Notably, the waiting time between the deployment and collection phases spans almost 10 years, with a duration of 3334 days.

\subsection{\label{sec:BIP}Binary Integer Programming}

BIP is used to solve the combinatorial ordering problem and provide guarantees on the optimality of the solution. The main challenge lies in finding an appropriate transcription of the problem that allows for efficient solving. These types of problems share a very similar structure to well-known routing problems, such as the traveling salesman \cite{lawlerTravelingSalesmanProblem1985} and optimal flow \cite{ahujaNetworkFlowsTheory1993} problems. Since both of these problems are commonly interpreted using a graph theory approach, we develop a similar framework to describe this formulation.

It is important to note that this formulation does not account for the initial and terminal transfers required for a valid GTOC 12 trajectory (i.e., the transfers from Earth to the asteroids and from the asteroids to Earth). Each subset contains a variety of asteroids with similar characteristics, meaning that these initial and terminal transfers would approximately have equal cost across different asteroids. Thus, their choice should only have minimal impact on the optimality of the trajectory. During our development and testing, this was tested with a modified formulation that included the cost of these transfers, but because no significant differences in solution quality were observed, they are omitted from the final formulations to simplify the problem. 

Using graph theory terminology, a node is created for each potential rendezvous target in our subset, while the transfers between them are represented by arcs. Unfortunately, as in many spaceflight optimization problems, the cost to transfer between targets can vary significantly over time and often in an unpredictable manner. This marks a significant difference from standard traveling salesman-like problems. To work around this limitation, the continuous transfer costs are translated into a fixed rendezvous time schedule. As a consequence, transfers in the BIP can only start and end at predetermined times, which makes the complete enumeration of all possible transfer arcs tractable. To account for this change, a separate set of nodes is introduced for each of the fixed rendezvous times, with directed arcs connecting them to represent the transfers between asteroids. The costs of these arcs are therefore fixed since the rendezvous times are predetermined.

In summary, starting from a set of possible rendezvous targets, each target is represented by a requisite node for each rendezvous time. Between each of these sets of nodes are directed arcs, which each represent a transfer between asteroids. The cost of each of these arcs is fixed due to the known time schedule at which each rendezvous occurs. Therefore, a natural BIP formulation would be to introduce a binary variable for each possible arc. 

\subsubsection{Self-cleaning ship formulation}

In the GTOC 12 problem, the deployment and collection rendezvous events need to be distinguished from one another. To achieve this within the BIP formulation, separate binary variables are introduced: $d_{i, j, k} \in [0, 1]$ for deployment events and $c_{i, j, k} \in [0, 1]$ for collection events. Additionally, the variable $m_{i, j} \in [0, 1]$ is defined to represent the intermediate transfer phase between the deployment and collection phases of a mission. In the notation for these variables, the first subscript ($i$) represents the departure target, the second subscript ($j$) is the arrival target, and the third subscript ($k$) denotes the stage in the rendezvous schedule. When a subscript index is demarked with \textit{end}, it indicates the selection of the last valid index for that specific subscript. 

\begin{figure}
    \centering
    \includegraphics[width = \textwidth]{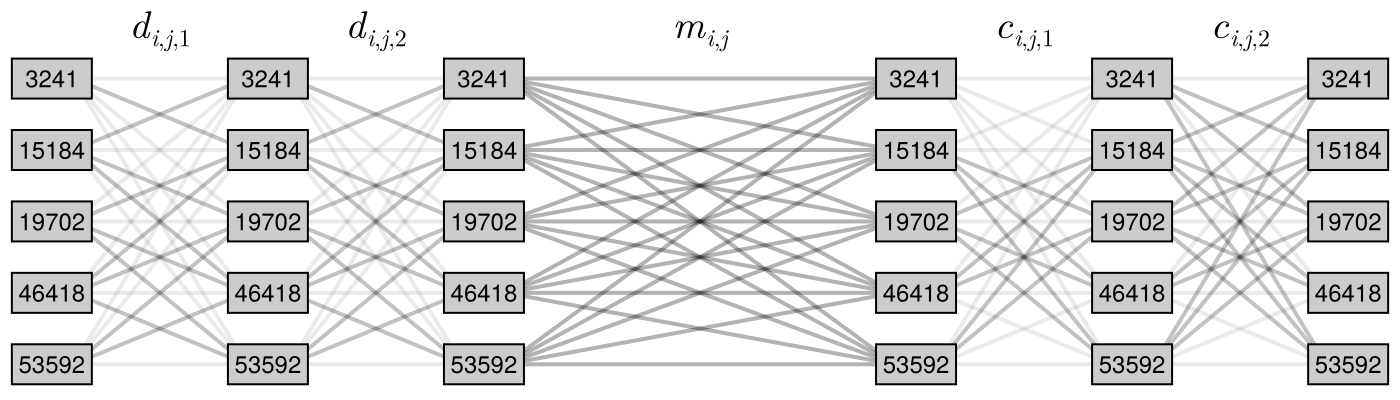}
    \caption{An example of the graph structure of the BIP variables for GTOC 12 problems.}
    \label{bip_connected}
\end{figure}

For example, to create the structure for the deployment and collection of 3 miners in a single mission from the example subset of size 5, $d_{i, j, k}$ and $c_{i, j, k}$ would have indices $i$ and $j$ from 1 to 5 (representing the choice of asteroid) and $k$ from 1 to 2 (representing the 2 directed arcs that connect 3 rendezvous events). Similarly, the $m_{i, j}$ would have $i$ and $j$ from 1 to 5. A visual representation of this graph structure is shown in Figure~\ref{bip_connected}.

The visualization of this graph structure immediately highlights the potential for optimizations by removing unnecessary arcs, corresponding to the elimination of variables within the BIP formulation. In order to help facilitate this process, the $d$, $m$, and $c$ variable arrays are initialized as sparse matrices, where only variables that meet certain criteria are included in the problem formulation. 

Firstly, arcs connecting to the same target during the deployment or collection phases of the mission are not feasible and can be removed, formally $d_{i, j, k}$ where $i=j$ and $c_{i, j, k}$ where $i=j$. However, for the intermediate variables $m_{i, j}$, these transfers are permitted since it is often optimal to wait at the same target between the deployment and collection phases. Additionally, a cost-based pruning step removes arcs that are unlikely to be used due to their high predicted costs. For the purposes of the GTOC 12 problem, this criterion was typically set to when the calculated Lambert $\Delta v$ of a specific transfer exceeded $6$-$10$ km/s. This criterion is applied across all arcs within the formulation. 

After applying these pruning steps, the resultant graph structure is illustrated by the bold arcs in Figure~\ref{bip_connected}. The total number of binary variables decreased from 125 in the original formulation to 61 in the pruned formulation. It is important to note that in some cases, this pruning can be too aggressive and lead to problem infeasibility; this can be easily addressed by increasing the cutoff for the $\Delta v$ threshold. 

Several constraints are required within the BIP formulation in order to produce feasible rendezvous sequences for the GTOC 12 problem (or even rendezvous sequence problems in general). The precise formulations of these constraints are very important, as they should only restrict transfers that are genuinely infeasible. Each of these constraints is detailed in the following paragraphs.

\textbf{(single arc selected per stage)} This constraint ensures that a trajectory contains only a single active arc per stage, which represents a single transfer event. Collectively, these constraints guarantee that across the whole trajectory, only one arc is selected at each stage:
\begin{align} \label{bip_single_per_stage}
    \forall k: \sum_{i, j} d_{i, j, k} = 1 && \sum_{i, j} m_{i, j} = 1 && \forall k: \sum_{i, j} c_{i, j, k} = 1
\end{align}

\textbf{(arc departs if an arc arrives)} If an arc arrives at a specific asteroid, another arc must depart from that same asteroid in the next stage of the trajectory. This ensures the continuity of the rendezvous sequence:
\begin{align} \label{bip_continuity}
    \forall j, \forall k: \sum_{i} d_{i, j, k} = \sum_{i} d_{j, i, k+1} && \forall j, \forall k: \sum_{i} c_{i, j, k} = \sum_{i} c_{j, i, k+1}
\end{align}

\textbf{(deployment/collection once per phase)} Deployments and collections must not occur more than once at the same asteroid. This constraint checks all outgoing arcs from each asteroid in each stage to prevent duplicates. Additionally, the incoming arcs for the last asteroids are considered, as they do not have outgoing arcs:
\begin{align} \label{bip_single_deploy_collect}
    \forall i: \sum_{j, k} d_{i, j, k} + \sum_{j} d_{j, i, \text{end}} \leq 1 && \forall i: \sum_{j, k} c_{i, j, k} + \sum_{j} c_{j, i, \text{end}} \leq 1
\end{align}

\textbf{(intermediate variable selection)} This constraint ensures that the intermediate variable assumes its correct value based on the selected final deployment and first collection. If a specific $m$ is selected, the corresponding $d$ and $c$ variables must also be activated due to the binary nature of the problem:
\begin{align} \label{bip_intermediate}
    \forall i, \forall j: 2m_{i, j} \leq \sum_{k} d_{k, i, \text{end}} + \sum_{k} c_{j, k, 1}
\end{align}

Alternatively, the intermediate variable can be omitted from the formulation, and instead replaced with an expression in terms of the final deployment and first collection variables. In our testing, this did not make a discernible difference in practice, likely due to the optimizer's resolve routines, and the replacement only tended to complicate the formulation. 

\textbf{(collection occurs if deployment)} Collection from an asteroid should only occur if a miner deployment has occurred. Although it is technically possible within the GTOC 12 problem to deploy a miner without collecting it, this scenario is suboptimal and should not arise in any high-quality solution. This relationship is expressed with the following equality constraint:
\begin{align} \label{bip_deploy_collect}
    \forall i: \sum_{j, k} d_{i, j, k} + \sum_{j} d_{j, i, end} = \sum_{j, k} c_{i, j, k} + \sum_{j} c_{j, i, \text{end}}
\end{align}

Together, these constraints ensure that the binary variables reflect a feasible self-cleaning solution for the GTOC 12 problem. The objective of the BIP is to minimize the total transfer cost associated with a feasible solution, expressed as:
\begin{align} \label{bip_objective}
    J = \sum_{i, j, k} \alpha_{i, j, k} d_{i, j, k} + \sum_{i, j} \beta_{i, j} m_{i, j} + \sum_{i, j, k} \gamma_{i, j, k} c_{i, j, k}
\end{align}
where $\alpha_{i, j, k}$ represents the cost for each deployment arc, $\beta_{i, j}$ is the cost of each intermediate arc and $\gamma_{i, j, k}$ is the cost of each collection arc. These costs correspond to the $\Delta v$ values derived from the lowest-cost solution to single- and multi-revolution Lambert problems for each arc. While higher-quality approximations for arc costs, such as those accounting for true low-thrust control, can be employed, they come with increased computational expense, which may end up dominating the cost of the BIP step. Therefore, the full BIP problem to solve can be framed as:
\begin{mini}[2]
    {d_{i, j, k}, m_{i, j, k}, c_{i, j, k}}{\eqref{bip_objective}}{}{}
    \addConstraint{\eqref{bip_single_per_stage}}{}{\quad\text{(single arc selected per stage)}}
    \addConstraint{\eqref{bip_continuity}}{}{\quad\text{(arc departs if an arc arrives)}}
    \addConstraint{\eqref{bip_single_deploy_collect}}{}{\quad\text{(deployment/collection once per phase)}}
    \addConstraint{\eqref{bip_intermediate}}{}{\quad\text{(intermediate variable selection)}}
    \addConstraint{\eqref{bip_deploy_collect}}{}{\quad\text{(collection occurs if deployment)}}
    \addConstraint{d_{i, j, k}, m_{i, j, k}, c_{i, j, k}}{\in \{0, 1\}}{\quad\text{(binary variables)}}
    \label{eq:convex_problem}
\end{mini}

This problem is formulated and solved using the \texttt{JuMP} \cite{lubinJuMPRecentImprovements2023} modeling language, which easily allows for the drop-in use of a variety of solvers supporting BIP problems, which are often referred to as more general mixed-integer programming solvers. Both the commercial Gurobi \cite{gurobioptimizationllcGurobiOptimizer2023} and open-source HiGHS \cite{huangfuParallelizingDualRevised2018} can be used to solve problems in this formulation efficiently. 

A key advantage of this formulation approach lies in the solving process. Solvers for these types of problems typically allow for not only the best solution to be extracted but also the set of the best solutions by the introduction of blocking constraints. This enables a more detailed evaluation step where the true objective function can be used to identify the most optimal solution, which in this case would involve solving the optimal control problem. This capability, stemming from the BIP solution process, enables a significant increase in the exploration of the objective landscape, resulting in improved solution quality.

\begin{figure}
    \centering
    \includegraphics[width = \textwidth]{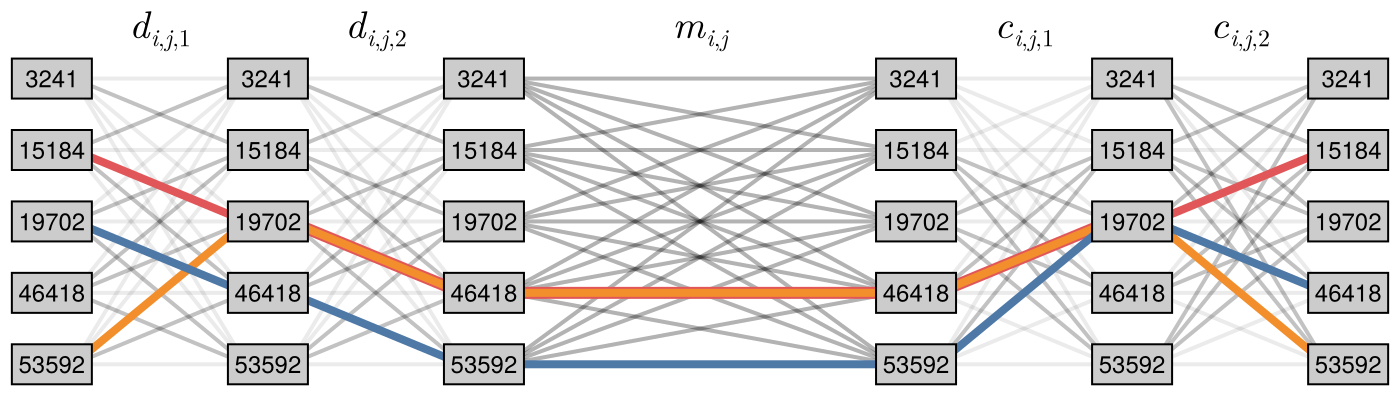}
    \caption{Top 3 solutions to the example BIP problem found in Figure~\ref{bip_connected} with 3 deployments and 3 collections.}
    \label{bip_solutions}
\end{figure}

\begin{table}
    \centering
    \caption{Details on the solutions found in Figure~\ref{bip_solutions}.}
    \begin{tabular}{lrrrrrr}\toprule
         & \multicolumn{2}{l}{\textbf{Rank 1}} & \multicolumn{2}{l}{\textbf{Rank 2}} & \multicolumn{2}{l}{\textbf{Rank 3}} \\
        \cmidrule(l){2-3} \cmidrule(l){4-5} \cmidrule(l){6-7}
        \textbf{Event} & \textbf{ID} & $\mathbf{\Delta v}$ & \textbf{ID} & $\mathbf{\Delta v}$ & \textbf{ID} & $\mathbf{\Delta v}$\\
        \midrule
        Miner Deployment 1 & 19702 & - & 53592 & - & 15184 & - \\
        Miner Deployment 2 & 46418 & 1.14 & 19702 & 4.78 & 19702 & 5.59 \\ 
        Miner Deployment 3 & 53592 & 4.22 & 46418 & 1.03 & 46418 & 1.03 \\
        Mass Collection 1 & 53592 & 0.00 & 46418 & 0.00 & 46418 & 0.00 \\
        Mass Collection 2 & 19702 & 3.98 & 19702 & 4.02 & 19702 & 4.02 \\ 
        Mass Collection 3 & 46418 & 3.52 & 53592 & 3.09 & 15184 & 2.89 \\
        \midrule
        \textbf{Total transfer} $\mathbf{\Delta v} \left[\textbf{km/s}\right]$ & & 12.86 & & 12.92 & & 13.53\\
        \bottomrule
    \end{tabular}
    \label{bip_solutions_table}
\end{table}

\begin{figure}
    \centering
    \includegraphics[width = \textwidth]{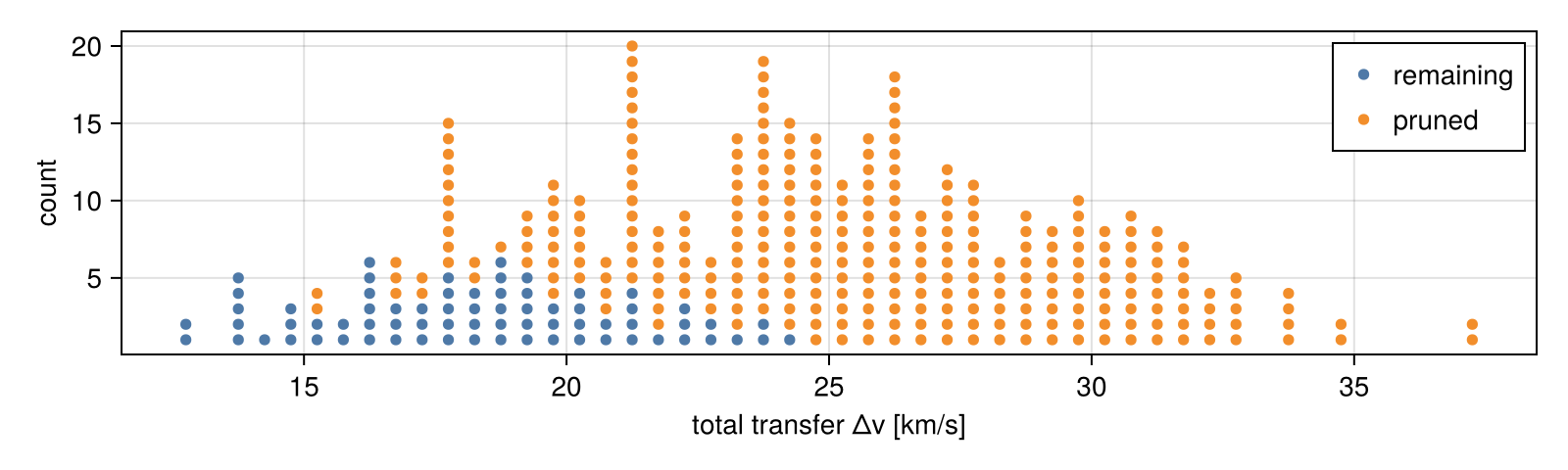}
    \caption{Objective landscape of the solution pool before and after pruning.}
    \label{bip_solution_values}
\end{figure}

An example of the solutions that are obtained through the BIP process is illustrated in Figure~\ref{bip_solutions}, where the top 3 solutions found to the example problem (outlined in Tables~\ref{restricted_parameters} and \ref{transfer_times_restricted}) are shown in blue, orange, and red respectively. More details on the differences between these solutions can be found in Table~\ref{bip_solutions_table}. Each of these solutions adheres to the constraints set within the GTOC 12 problem, and each demonstrates taking a slightly different path through the available rendezvous targets, resulting in slightly different transfer costs.

Additionally, Figure~\ref{bip_solution_values} demonstrates the impact of the pruning step within the  BIP formulation, showing how it can effectively reduce the number of variables without compromising the quality of the best solutions for the example problem. In this case, the Lambert pruning cutoff is set to $6$ km/s. Before pruning, solving the BIP identifies a total of 360 possible solutions; after the pruning step, only 70 remain. However, of the solutions with high quality (those with low total transfer $\Delta v$), almost all are retained.

\subsubsection{Mixed ship formulation}

A clear limitation of the self-cleaning BIP formulation is its inability to optimize a critical aspect of the GTOC 12 problem: collaboration among different mining ships to develop a mixed strategy. Many of the highest-scoring solutions in the competition, including the winning strategy by JPL, heavily leveraged this to produce better-quality campaigns. To address this, several modifications to the self-cleaning BIP formulation are introduced to enable the optimization of mixed solutions using a similar methodology.

This process involves duplicating the self-cleaning formulation for each mixed ship. Rather than treating these as separate BIP problems, all variables and constraints are integrated into a single BIP and solved simultaneously. To account for the additional sets of variables required, the variables pertaining to each of these graph structures are denoted with a superscript index, such as $d^{\,l}_{i, j, k}$. Then, some of the constraints for each mixed ship need to be modified in order to generate feasible solutions. 

Firstly, while the constraints for selecting a single arc per stage \eqref{bip_single_per_stage}, making sure an arc departs if one arrives \eqref{bip_continuity}, and selecting the correct intermediate variable \eqref{bip_intermediate} are unchanged, they now apply to each ship individually (i.e., one copy of each constraint for each superscript index). Then, some of the other constraints need to be further altered, which is detailed in the following paragraphs.

\textbf{(deployment/collection once per phase)} Since deployment and collection events can now occur with different ships, the constraint ensuring only a single deployment and collection at each asteroid occurs \eqref{bip_single_deploy_collect} must be modified to allow different ships to perform each operation separately:
\begin{align} \label{bip_multi_deploy_collect}
    \forall i: \sum_{j, k, l} d^{\,l}_{i, j, k} + \sum_{j, l} d^{\,l}_{j, i, \text{end}} \leq 1 && \forall i: \sum_{j, k, l} c^{\,l}_{i, j, k} + \sum_{j, l} c^{\,l}_{j, i, \text{end}} \leq 1
\end{align}

\textbf{(collection occurs if deployment)} To ensure that a collection event occurs only if a deployment is made, constraint \eqref{bip_deploy_collect} must also be modified similarly:
\begin{align} \label{bip_deploy_collect_multi}
    \forall i: \sum_{j, k, l} d^{\,l}_{i, j, k} + \sum_{j, l} d^{\,l}_{j, i, \text{end}} = \sum_{j, k, l} c^{\,l}_{i, j, k} + \sum_{j, l} c^{\,l}_{j, i, \text{end}}
\end{align}

Notably, the collaboration between multiple ships allows for the possibility of ship trajectories that make differing numbers of deployments and collections. The only restriction is that the total number of deployment events must equal the total number of collection events. This fits naturally into the formulation; the only change is that the third index for deployment and collection on each ship may no longer increment to the same value. Care must be taken that the intermediate variable selection constraint \eqref{bip_intermediate} is correctly implemented, considering that the \textit{end} index may vary between ships.

Finally, the objective for the mixed ship BIP formulation becomes:
\begin{align} \label{bip_objective_multi}
    J = \sum_{i, j, k, l} \alpha^{\,l}_{i, j, k} d^{\,l}_{i, j, k} + \sum_{i, j, l} \beta^{\,l}_{i, j} m^{\,l}_{i, j} + \sum_{i, j, k, l} \gamma^{\,l}_{i, j, k} c^{\,l}_{i, j, k}
\end{align}

In this formulation, instead of solely considering the cost of a single self-cleaning ship trajectory, the goal is to minimize the combined cost of all trajectories. The cost variables $\alpha$, $\beta$, and $\gamma$ are additionally assigned a superscript index for two main reasons. First, the lengths of each ship's trajectory may differ, necessitating distinct sizes of transfer costs. Second, the fixed rendezvous time schedules are likely to vary between ships, resulting in unique transfer costs that cannot be shared. 

\begin{figure}
    \centering
    \includegraphics[width = \textwidth]{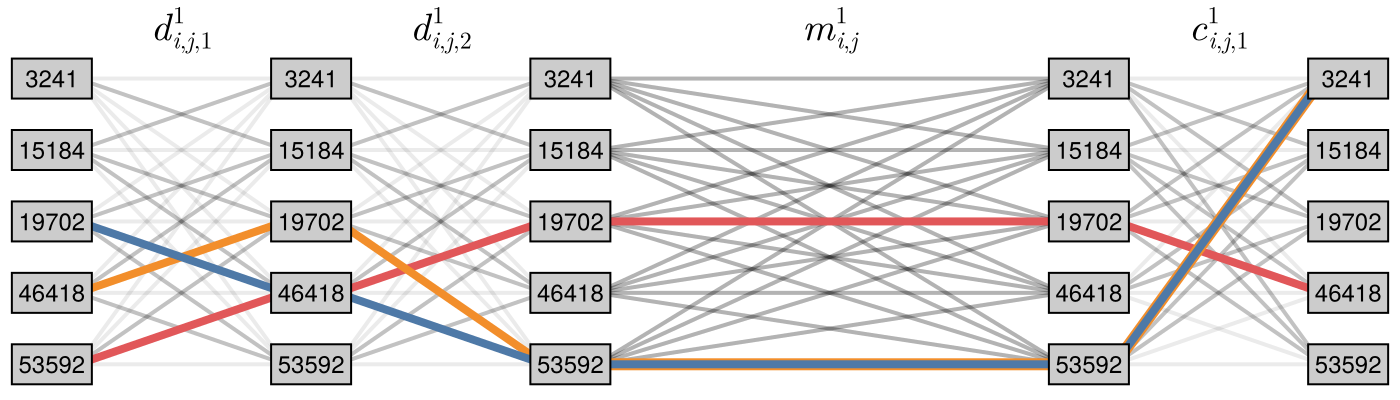}
    \\[3mm]
    \includegraphics[width = \textwidth]{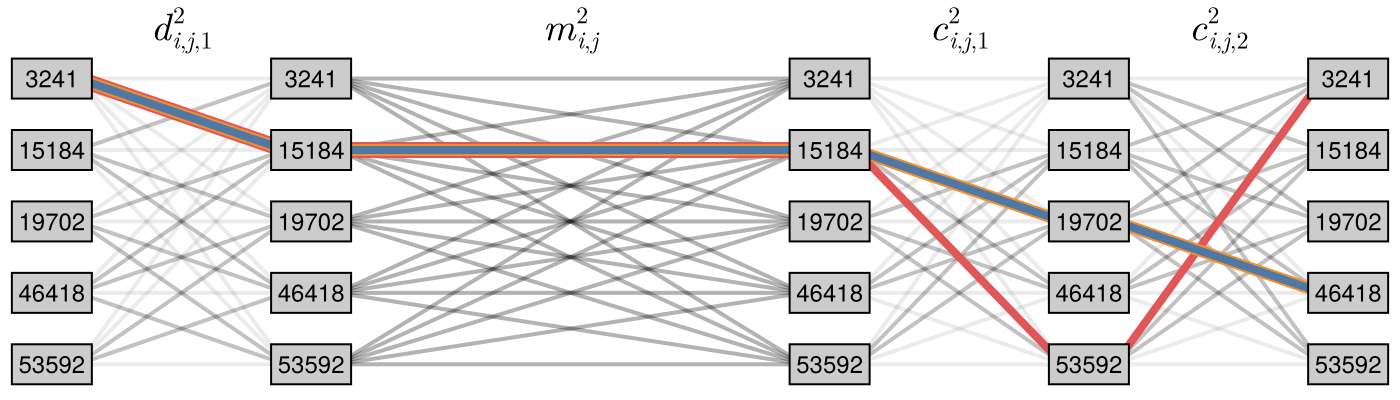}
    \caption{Top 3 solutions to the example BIP problem found in Figure~\ref{bip_connected} with 2 mixed ships.}
    \label{bip_solutions_mixed}
\end{figure}

Figure~\ref{bip_solutions_mixed} demonstrates the graph structure and the top 3 solutions of the BIP for a campaign involving 2 ships (in blue, orange and red respectively). One ship has 3 deployment events and 2 collection events, whilst the other has 2 deployment events and 3 collection events. Clearly, this modified BIP formulation can accommodate the optimization of mixed ships with differing amounts of deployments than collections. Similarly to the self-cleaning formulation, the extraction of multiple solutions can be directly applied to this case in order to improve the search behavior.

Additionally, this formulation can be used to simultaneously optimize sets of self-cleaning ships by reincorporating constraint \eqref{bip_single_deploy_collect}. This permits only a single deployment and collection to be made at each asteroid, and that each ship must be self-cleaning. This is particularly important when searching for sets of self-cleaning solutions in subsets that may have multiple high-quality solutions, ensuring they do not overlap in terms of the asteroids visited. 

\subsection{\label{sec:SCP}Sequential Convex Programming}
To solve for the optimal control profile and the optimal rendezvous time schedule, an adaptive-mesh SCP is used \cite{kumagaiAdaptiveMeshSequentialConvex2024}. In contrast to typical implementations of SCP for spacecraft trajectory optimization problems, this method incorporates the time variables into the optimization process, enabling the mesh discretization to be optimized simultaneously with the trajectory. To accommodate the highly multi-target trajectories that are present in the GTOC 12 problem, these methods have been extended to support the simultaneous optimization of multiple rendezvous events throughout the trajectory. It is then further extended to support the optimization of multiple ship trajectories simultaneously in a manner similar to the BIP. Firstly, the formulation of the SCP problem for a single ship trajectory is introduced.

\subsubsection{Single trajectory optimization}

Each individual ship trajectory consists of multiple rendezvous events. In the GTOC 12 problem, some of these events involve the deployment of miners, while others pertain to the collection of mined mass. Between each successive event, there exists a trajectory arc on which the optimal control profile must be determined; this is referred to as a {\it leg} of a trajectory. Each leg is discretized into many individual parts over time as in a direct method, and these parts are referred to as {\it segments}. For the GTOC 12 problem, the number of segments in a leg is determined based on a predefined segment timespan, which is chosen prior to the optimization. For instance, a 201-day leg with 5-day segments would have a total of 40 5-day segments culminating with a terminal 1-day segment. 

Similar to the subscripts used in the BIP formulation, the first subscript for each variable denotes the segment, while the second subscript indicates the leg of the trajectory. When a subscript index is marked with \textit{end}, this signifies taking the last valid index for that specific subscript.

Due to the incorporation of the time variable into the adaptive-mesh SCP, this discretization primarily serves as an initial guide for the optimization process because the timespan of each segment is selected during the optimization process. This parameter acts as a scaling factor for the timespan of each individual segment and is optimized during the SCP process. Consequently, the exact timespans of the segments may vary between segments, allowing the optimization to adapt the trajectory segmentation and potentially identify better quality solutions. 

\begin{figure}
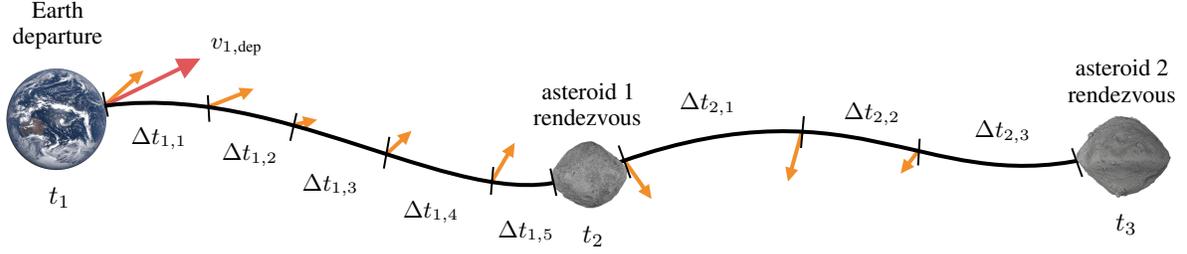

    \centering
    \include{figures/scp_explainer}
    \caption{Underlying structure of an optimized SCP problem for the first 2 legs of a problem.}
    \label{scp_explainer}
\end{figure}

As a result, without a fixed integrality constraint across all segments, it becomes evident that the rendezvous times could change to enhance optimality. This flexibility, which is enabled by including the segment timespans into the optimization, is crucial to obtaining solutions that are optimal not only in terms of control profile but also regarding their choice of rendezvous time schedule. Figure~\ref{scp_explainer} demonstrates the underlying structure of the adaptive-mesh SCP formulation, showing the time discretization with black segmentation, and the control vectors in orange.

Applying the general approach of SCP \cite{malyutaConvexOptimizationTrajectory2022}, the process begins by finding an appropriate linearization of the system dynamics for each trajectory segment. The Keplerian dynamics of the mining ship are provided in Equation~\eqref{dynamical_equation} with the state vector $\mathbf{x}=[\mathbf{r}, \mathbf{v}, m]$ and control vector as $\mathbf{u}=[\mathbf{T}]$. This formulation of the dynamics is referred to as {\it true-mass}. 

Alternatively, an expression of the dynamics working with the logarithm of the mass \cite{acikmeseConvexProgrammingApproach2007} is denoted {\it logged-mass}. This formulation tended to exhibit superior convergence characteristics when compared to the true-mass formulation. The change of variables is as follows:
\begin{align}
    \mathbf{\Gamma} = \frac{\mathbf{T}}{m} && w = \ln m
\end{align}

This leads to the dynamical equations with the adjusted state vector $\mathbf{x}=[\mathbf{r}, \mathbf{v}, w]$ and control vector $\mathbf{u}=[\mathbf{\Gamma}]$:
\begin{align}
    \dot{\mathbf{x}} = f(\mathbf{x}, \mathbf{u}) = \left\{\begin{array}{l}
            \dot{\mathbf{r}} = \mathbf{v} \\
            \dot{\mathbf{v}} = -\frac{\mu}{r^3} \mathbf{r} + T_{\text{max}} \mathbf{\Gamma} \\
            \dot{w} = -\frac{T_{\text{max}}}{I_{\text{sp}} g_0} \Gamma
        \end{array}\right.
\end{align}
where $\mu$ is the gravitational constant of the Sun, $w$ represents the {\it logged} spacecraft mass, and $0 \leq \Gamma \leq e^{-w}$ is the new control variable representing the spacecraft thrust divided by mass, which is normalized to equal $T_{\text{max}}$ when $m=m_{\text{max}}$. This is the formulation primarily used throughout the SCP process.

In both formulations, the control vector $\mathbf{u}$ is kept constant across a segment, corresponding to a zero-order hold (ZOH) discretization. In the subsequent paragraphs, the constraints required for the SCP program are introduced, along with the underlying variables used in the formulation.

\textbf{(linearized dynamics)} The linearized dynamic constraints are constructed around a reference trajectory. The SCP process requires an appropriate initial guess for the reference trajectory, which should be close in position to the optimal low-thrust trajectory, but it does not necessarily need to be dynamically feasible. For the case of GTOC 12, single- or multi-revolution Lambert arcs are used, which worked well for the reference trajectory. Then, in the subsequent iterations of the SCP algorithm, the control profile found at each previous iteration can be employed to generate a reference trajectory. Given a reference trajectory, thrust profile, and segment time $(\mathbf{\bar{x}}, \mathbf{\bar{u}}, \bar{\Delta t})$, a convex constraint is constructed based on the dynamical linearization of each segment:
\begin{align} \label{dynamics_linear}
    \forall i, \forall j: \mathbf{x}_{i, j+1} = \mathbf{A}_{i, j}\mathbf{x}_{i, j} + \mathbf{B}_{i, j}\mathbf{u}_{i, j} + \mathbf{C}_{i, j} \Delta t_{i, j} + \mathbf{d}_{i, j}
\end{align}

Here, $i$ represents the leg index and $j$ the segment index. The dynamical linearization matrices are computed as follows:
\begin{equation}
\begin{aligned}
    \mathbf{A}_{i, j} =& \left. \frac{\partial}{\partial \mathbf{x}} \int_{0}^{\Delta t}\dot{\mathbf{x}}\,\text{d}\tau\;\right|_{(\mathbf{\bar{x}}_{i, j}, \mathbf{\bar{u}}_{i, j}, \bar{\Delta t}_{i, j})} && 
    \mathbf{B}_{i, j} = \left. \frac{\partial}{\partial \mathbf{u}} \int_{0}^{\Delta t} \dot{\mathbf{x}}\,\text{d}\tau\; \right|_{(\mathbf{\bar{x}}_{i, j}, \mathbf{\bar{u}}_{i, j}, \bar{\Delta t}_{i, j})} \\ 
    \mathbf{C}_{i, j} =& \left. \frac{\partial}{\partial \Delta t} \int_{0}^{\Delta t} \dot{\mathbf{x}}\,\text{d}\tau\; \right|_{(\mathbf{\bar{x}}_{i, j}, \mathbf{\bar{u}}_{i, j}, \bar{\Delta t}_{i, j})} &&
    \mathbf{d}_{i, j} = \mathbf{\bar{x}}_{i, j+1} - \mathbf{A}_{i, j} \mathbf{\bar{x}}_{i, j} - \mathbf{B}_{i, j} \mathbf{\bar{u}}_{i, j} - \mathbf{C}_{i, j} \bar{\Delta t}_{i, j}
\end{aligned}
\end{equation}

Thus, the discrete form of the dynamics is obtained for each segment $j$ of each leg $i$, which can be used as a convex constraint. The matrix $\mathbf{A}_{i, j}$ is the state transition matrix (STM), representing the changes in the final state with respect to changes in the initial state. Correspondingly, $\mathbf{B}_{i, j}$ represents the changes in the final state with respect to the control, $\mathbf{C}_{i, j}$ the changes in the final state with respect to the segment duration, and $\mathbf{d}_{i, j}$ is the offset vector for the linearization.

These partial derivatives are computed via automatic differentiation (AD) directly applied to the initial conditions of a numerical integration solver which performs the integration. The Vern9 \cite{vernerExplicitRungeKutta1978} numerical integrator is used from the \texttt{DifferentialEquations.jl} library \cite{rackauckasDifferentialEquationsJlPerformant2017} with absolute tolerance $10^{-12}$ and relative tolerance $10^{-12}$. The AD is computed through the use of \texttt{ForwardDiff.jl} \cite{revelsForwardModeAutomaticDifferentiation2016}. 

\textbf{(control magnitude)} For the control magnitude constraint, a convex second-order-cone (SOC) constraint is used to obtain the $L_2$ norm of the control vector. The control magnitude $u_{i, j}$ is an additional variable within the optimization process. This constraint will bind at optimality because the objective function either directly or indirectly minimizes control usage:
\begin{align} \label{control_magnitude}
    \forall i, \forall j: u_{i, j} \geq \|\mathbf{u}_{i, j}\|_2
\end{align}

\textbf{(control limit)} Enforcing the control limits varies depending on the mass formulation used. The logged-mass variant is more complex because it needs an additional linearization step, as the maximum control magnitude changes based on the mass. The Taylor expansion of $\Gamma = e^{-w} T$ is taken around a reference logged-mass $\bar{w}$ to establish an appropriate control limit for the constraint:
\begin{equation} \label{control_limit}
\begin{aligned}
    (\text{true-mass})\, \forall i, \forall j &: u_{i, j} \leq 1 \\
    (\text{logged-mass})\, \forall i, \forall j &: u_{i, j} \leq e^{-\bar{w}_{i, j}} (1 - w_{i, j} + \bar{w}_{i, j})
\end{aligned}
\end{equation}

\textbf{(linearized rendezvous)} The initial and final state constraints on each leg of the trajectory must also be linearized. For simplicity, an expression for the time in terms of other known variables (notably $\Delta t_{i, j}$) is introduced:
\begin{align} \label{actual_time}
    t_{i + 1} = \sum_{j} \Delta t_{i, j} + t_{i}
\end{align}
where $t_i$ is an expression representing the actual start time of leg $i$. An additional variable for $t_1$ needs to be introduced representing the start time. For each segment, the initial and terminal state constraints are introduced as:
\begin{align} \label{rendezvous_dynamics_linear_start}
    \forall i:& \left[\begin{array}{c}
        \mathbf{r}_{i, 1} \\ \mathbf{v}_{i, 1}
    \end{array} \right] = \mathbf{e}_i t_i + \mathbf{f}_i + \left[\begin{array}{c}
        \mathbf{0} \\ \Delta \mathbf{v}_{i, \text{dep}}
    \end{array} \right] + \bm{\delta}_{i, \text{dep}}
    \\ \label{rendezvous_dynamics_linear_end}
    \forall i:& \left[\begin{array}{c}
        \mathbf{r}_{i, \text{end}} \\ \mathbf{v}_{i, \text{end}}
    \end{array} \right] = \mathbf{e}_{i+1} t_{i+1} + \mathbf{f}_{i+1} + \left[\begin{array}{c}
        \mathbf{0} \\ \Delta \mathbf{v}_{i, \text{arr}}
    \end{array} \right] + \bm{\delta}_{i, \text{arr}}
\end{align}
In these equations, for leg $i$, $\Delta \mathbf{v}_{i, \text{dep}}$ represents additional velocity added on departure, while $\Delta \mathbf{v}_{i, \text{arr}}$ represents additional velocity added on arrival. The variables $\bm{\delta}_{i, \text{dep}}$ and $\bm{\delta}_{i, \text{arr}}$ are slack variables representing the rendezvous constraint violation at departure and arrival, respectively. The vectors $\mathbf{e}$ and $\mathbf{f}$ are expressed as: 
\begin{equation}
\begin{aligned}
    \mathbf{e}_i = \left. \frac{\partial \mathbf{x}_i}{\partial t}
    \;\right|_{(\bar{t}_i)} &&
    \mathbf{f}_i = \mathbf{\bar{x}}_i - \mathbf{e}_i \bar{t}_i
\end{aligned}
\end{equation}
where for rendezvous target $i$, $\mathbf{x}_i$ represents the target state, $\bar{\mathbf{x}}_i$ the reference target state and $\bar{t}_i$ indicates the reference rendezvous time. The partial derivative with respect to time can be calculated using a variety of methods; in this case, AD is applied to the position ephemeris function. For GTOC 12, this function computes Keplerian dynamics based on classical orbital elements.

\textbf{(violation absolute value)} Additional variables $\bm{\delta}_{i, \text{dep}}^+$ and $\bm{\delta}_{i, \text{arr}}^+$, representing the absolute violation of the leg start and end constraints, are introduced. The following inequality constraints ensure that the absolute values are captured:
\begin{align} 
    \label{state_violation}
    \forall i: \, \bm{\delta}_{i, \text{dep}}^+ \geq \bm{\delta}_{i, \text{dep}} &&
    \forall i: \, \bm{\delta}_{i, \text{dep}}^+ \geq -\bm{\delta}_{i, \text{dep}} &&
    \forall i: \, \bm{\delta}_{i, \text{arr}}^+ \geq \bm{\delta}_{i, \text{arr}} &&
    \forall i:\, \bm{\delta}_{i, \text{arr}}^+ \geq -\bm{\delta}_{i, \text{arr}} 
\end{align}

\textbf{(departure and arrival $\Delta v$ limits)} Control limits also need to be applied to the additional velocities that could be permitted at the start and end phases of some legs:
\begin{align} 
    \label{dv_limits}
    \forall i: \, \Delta v_{i, \text{dep}} \geq \|\Delta \mathbf{v}_{i, \text{dep}}\|_2 && \forall i: \, \Delta v_{i, \text{arr}} \geq \|\Delta \mathbf{v}_{i, \text{arr}}\|_2
\end{align}
where $\Delta v_{i, \text{dep}}$ and $\Delta v_{i, \text{arr}}$ are selected based on the problem specification. For a typical trajectory in GTOC 12, without gravity assists, $\Delta v_{i, \text{dep}}$ is zero for all legs except for the first leg from Earth, where it is 6 km/s. Similarly, $\Delta v_{i, \text{arr}}$ is zero for all legs except for the terminal leg to Earth, where it is also set to 6 km/s.

\textbf{(initial and terminal time limits)} The minimum and maximum time of the trajectory is constrained, where $t_{\text{min}}$ and $t_{\text{max}}$ are defined based on the problem specification:
\begin{align} 
    \label{time_limits}
    t_1 \geq t_{\text{min}} && 
    t_{\text{end}} \leq t_{\text{max}}
\end{align}

\textbf{(mass linkage constraint)} Linkage constraints are introduced for the mass between legs, where the mass difference $\Delta m$ between two subsequent legs is calculated based on Equation~\ref{mined_mass}:
\begin{equation} \label{mass_deltas}
\begin{aligned} 
    \Delta m_i = \left\{\begin{array}{ll}
            -m_{\text{miner}} & t_{i} \in \text{deployment}\\
            m_{\text{rate}} \left( t_{i} - t_{\text{deployment}(i)} \right) & t_{i} \in \text{collection}
    \end{array}\right.
\end{aligned}
\end{equation}
where $t_{\text{deployment}(i)}$ represents the time at which the miner deployment occurs at the asteroid visited at the start of leg $i$. This expression is used directly in the true-mass formulation but cannot be used in the logged-mass formulation. Instead, a reference $\Delta \bar{m}_{i}$ is calculated based on the rendezvous times of the previous iteration of SCP. Then, the mass linkage constraint is expressed as:
\begin{equation} \label{mass_linkage}
\begin{aligned} 
    (\text{true-mass})\, \forall i, i>1:& \, m_{1, i} = m_{\text{end}, i-1} + \Delta m_{i} \\ 
    (\text{logged-mass})\, \forall i, i>1:& \, w_{1, i} = \log \left( e^{\bar{w}_{\text{end}, i-1}} + \Delta \bar{m}_{i} \right) + \frac{\left( w_{\text{end}, i-1} -  \bar{w}_{\text{end}, i-1} \right)}{1 + \Delta \bar{m}_{i} / e^{\bar{w}_{\text{end}, i-1}}} 
\end{aligned}
\end{equation}

The additional terms in the expression for the logged-mass constraint arise from the linearization of $\log \left( e^{w_{\text{end}, i-1}} + \Delta m_{i} \right)$ around a reference $\bar{w}_{\text{end}, i-1}$ from the previous iteration of SCP. 

\textbf{(initial and terminal mass)} The initial and terminal mass of the trajectory are constrained based on the problem specifications where $\gamma \ge 0$ represents a terminal mass violation, $m_\text{max}$ is the maximum (wet) mass of the ship, and $m_{\text{min}}$ is the minimum (dry) mass:
\begin{equation} \label{mass_limits}
\begin{aligned}
    (\text{true-mass})&: m_{1, 1} \leq m_\text{max}, \quad  m_{\text{end}, \text{end}} + \gamma \geq m_{\text{min}} + \sum_{i, \Delta \bar{m}_i > 0} \Delta m_{i} \\
    (\text{logged-mass})&: w_{1, 1} \leq \log(m_\text{max}), \quad w_{\text{end}, \text{end}} + \gamma \geq \log\left(m_{\text{min}} + \sum_{i, \Delta \bar{m}_i > 0} \Delta \bar{m}_{i}\right)
\end{aligned}
\end{equation}
Particularly, the terminal mass constraints additionally need to consider the final dropoff on the terminal leg during the GTOC 12 problem, where all the mined mass is deposited back at Earth. 

\textbf{(state and time trust regions)} In our previous implementations of SCP for GTOC 12 \cite{armellinGTOC12Results2024}, a trust region constraint was not required because the dynamics were sufficiently linear, as the time variable was not included in the optimization. This allowed the dynamical linearization around the reference trajectory to be sufficiently accurate for direct convergence within a few iterations. However, when compared to the other state variables, the introduction of the time variable presents a significant challenge to the dynamical linearization process because of its increased levels of nonlinear response. Therefore, a trust region constraint is advantageous when the time variable is included in the SCP. 

Trust regions are introduced as simple trust regions that decrease linearly in size as the SCP algorithm progresses. The trust region constraints for the state and time are separated and take the following form:
\begin{align} \label{trust_regions}
    \forall {i}, \forall {j}: -\epsilon_1 \leq \left[\begin{array}{c}
        \mathbf{r}_{i, j} - \mathbf{\bar{r}}_{i, j} \\ \mathbf{v}_{i, j} - \mathbf{\bar{v}}_{i, j} 
    \end{array} \right] \leq \epsilon_1 && \forall {i}, \forall {j}: -\epsilon_2 \leq \Delta t_{i, j} / \bar{\Delta t}_{i, j} - 1 \leq \epsilon_2
\end{align}
A variety of values for the initial size of the trust regions were tested, but values of $\epsilon_1=2.5$ and $\epsilon_2=0.05$ tended to perform well for the problems tested. These values are set to decrease linearly to zero over 80 iterations of the SCP algorithm, which helps to dampen oscillations during the convergence process. Improved variations of the trust region algorithms, such as state-dependent size based on nonlinearity indicators \cite{bernardiniStatedependentTrustRegion2024}, may provide superior performance, but this simple process worked effectively for the problem at hand. 

The objective for the SCP is to maximize the spread of the deployment and collection times, with penalties on leg start and end violations and penalties on the terminal mass violation:
\begin{align} \label{scp_objective}
    J = \sum_{i, t_{i} \in \text{deployment}} t_{i} - \sum_{i, t_{i} \in \text{collection}} t_{i} - 10^4 \sum_i \bm{\delta}_{i, \text{dep}}^+ - 10^4 \sum_i \bm{\delta}_{i, \text{arr}}^+ - 5\times 10^3 \gamma
\end{align}

Thus, the complete convex program solved at each iteration of SCP is:
\begin{maxi}[2]
    {\substack{
        \mathbf{x}, 
        \mathbf{u}, 
        u,
        \Delta \mathbf{t},
        \Delta \mathbf{v_{dep}}, 
        \Delta \mathbf{v_{arr}}, \\
        t_1,
        \gamma, 
        \bm{\delta}_{i, \text{dep}},
        \bm{\delta}_{i, \text{arr}},
        \bm{\delta}_{i, \text{dep}}^+,
        \bm{\delta}_{i, \text{arr}}^+
    }}{\eqref{scp_objective}}{}{}
    \addConstraint{\eqref{dynamics_linear}}{}{\quad(\text{linearized dynamics})}
    \addConstraint{\eqref{control_magnitude}, \eqref{control_limit}}{}{\quad(\text{control magnitude and control limit})}
    \addConstraint{\eqref{rendezvous_dynamics_linear_start}, \eqref{rendezvous_dynamics_linear_end}}{}{\quad(\text{linearized rendezvous})}
    \addConstraint{\eqref{state_violation}}{}{\quad(\text{violation absolute value})}
    \addConstraint{\eqref{dv_limits}}{}{\quad(\text{departure and arrival $\Delta v$ limits})}
    \addConstraint{\eqref{time_limits}}{}{\quad(\text{initial and terminal time limits})}
    \addConstraint{\eqref{mass_linkage}}{}{\quad(\text{mass linkage constraint})}
    \addConstraint{\eqref{mass_limits}}{}{\quad(\text{initial and terminal mass})}
    \addConstraint{\eqref{trust_regions}}{}{\quad(\text{state and time trust region})}
    \label{convex_problem}
\end{maxi}

The SCP process repeatedly solves the convex problem~\eqref{convex_problem} and updates the constraints \eqref{dynamics_linear}, \eqref{control_limit}, \eqref{rendezvous_dynamics_linear_start}, \eqref{rendezvous_dynamics_linear_end}, \eqref{mass_linkage} and \eqref{trust_regions} based on the reference trajectory obtained from the optimal solution of the previous iteration. The convergence of the SCP is tested based on two criteria: first, ensuring that the violation variables in the objective function are sufficiently small, and second, verifying the accuracy of the dynamical linearization against that of the full dynamical model. Both of these criteria are tested with a convergence parameter, which throughout our analysis was typically set to values between $10^{-6}$ and $10^{-8}$. 

Once convergence is achieved, the SCP yields the optimal control profile and the corresponding optimal rendezvous time schedule. To showcase the details of this process and the benefits of including the rendezvous time in the optimization, the optimal combinatorial ordering is used from the example problem, calculated via the BIP process in Figure~\ref{bip_solutions}. 

\begin{figure}
    \centering
    \includegraphics[width = 0.62\textwidth]{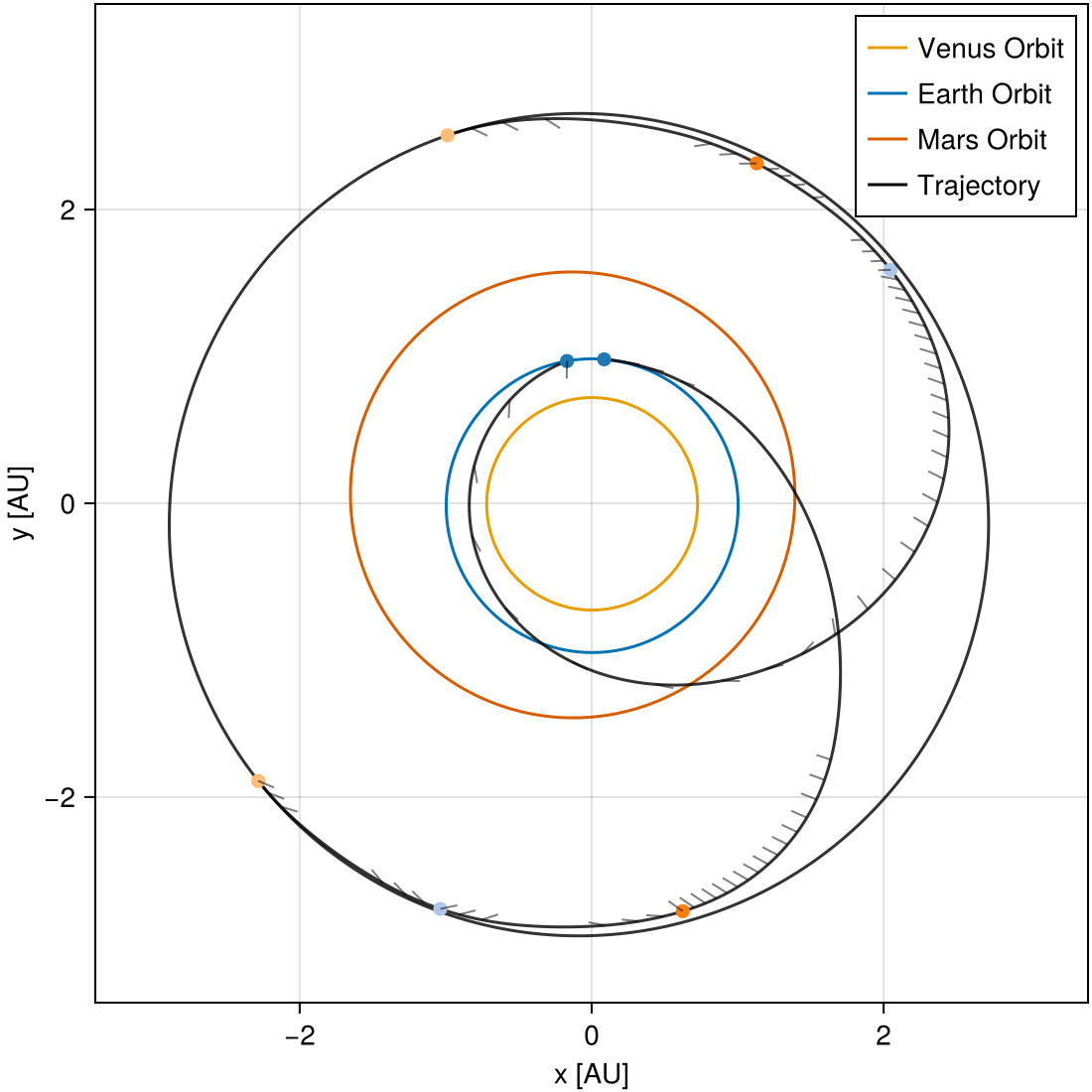}
    \caption{Optimal trajectory found using the SCP process for the combinatorial ordering found in Figure~\ref{bip_solutions}.}
    \label{scp_single_trajectory}
\end{figure}

\begin{table}
    \centering
    \caption{Rendezvous time schedule for Figure~\ref{scp_single_trajectory}.}
    \begin{tabular}{lrr}\toprule
        \textbf{Event} & \textbf{Time} $\left[\textbf{MJD}\right]$ & $\mathbf{\Delta t} \left[\textbf{days}\right]$\\
        \midrule
        Earth Departure & $64328.00$ & - \\
        Miner Deployment 1 & $64848.95$ & $520.95$\\
        Miner Deployment 2 & $64952.82$ & $103.87$\\ 
        Miner Deployment 3 & $65137.31$ & $184.49$\\
        Mass Collection 1 & $69109.33$ & $3972.02$\\
        Mass Collection 2 & $69256.70$ & $147.37$ \\ 
        Mass Collection 3 & $69413.11$ & $156.41$\\
        Earth Arrival & $69791.29$ & $378.18$\\
        \bottomrule
    \end{tabular}
    \label{transfer_times_single}
\end{table}

Figure~\ref{scp_single_trajectory} demonstrates the resultant optimal trajectory from the SCP process, with the corresponding optimal rendezvous times detailed in Table~\ref{transfer_times_single}. It is apparent that the SCP process has significantly adjusted the rendezvous times with the goal of increasing the returned mass. Specifically, all deployment events were shifted earlier, while collection events were delayed, extending the time between the rendezvous events at each asteroid. This can be observed when comparing Table~\ref{transfer_times_restricted} and Table~\ref{transfer_times_single}, where the intermediate transfer duration between deployment and collection phases increased from 3334.00 to 3927.02 days.

Interestingly, the mass return of this trajectory is not primarily limited by the maximum mass constraint \eqref{mass_limits} as may be expected. Instead, the mass is constrained by the minimum departure time. In this case, the optimal departure time of the spacecraft was found to be 64238 MJD - the earliest possible - indicating that the initial time constraint \eqref{time_limits} is binding. This represents a forward shift in start time of approximately 110 days from the initial rendezvous time guess.

\begin{figure}
    \centering
    \includegraphics[width = \textwidth]{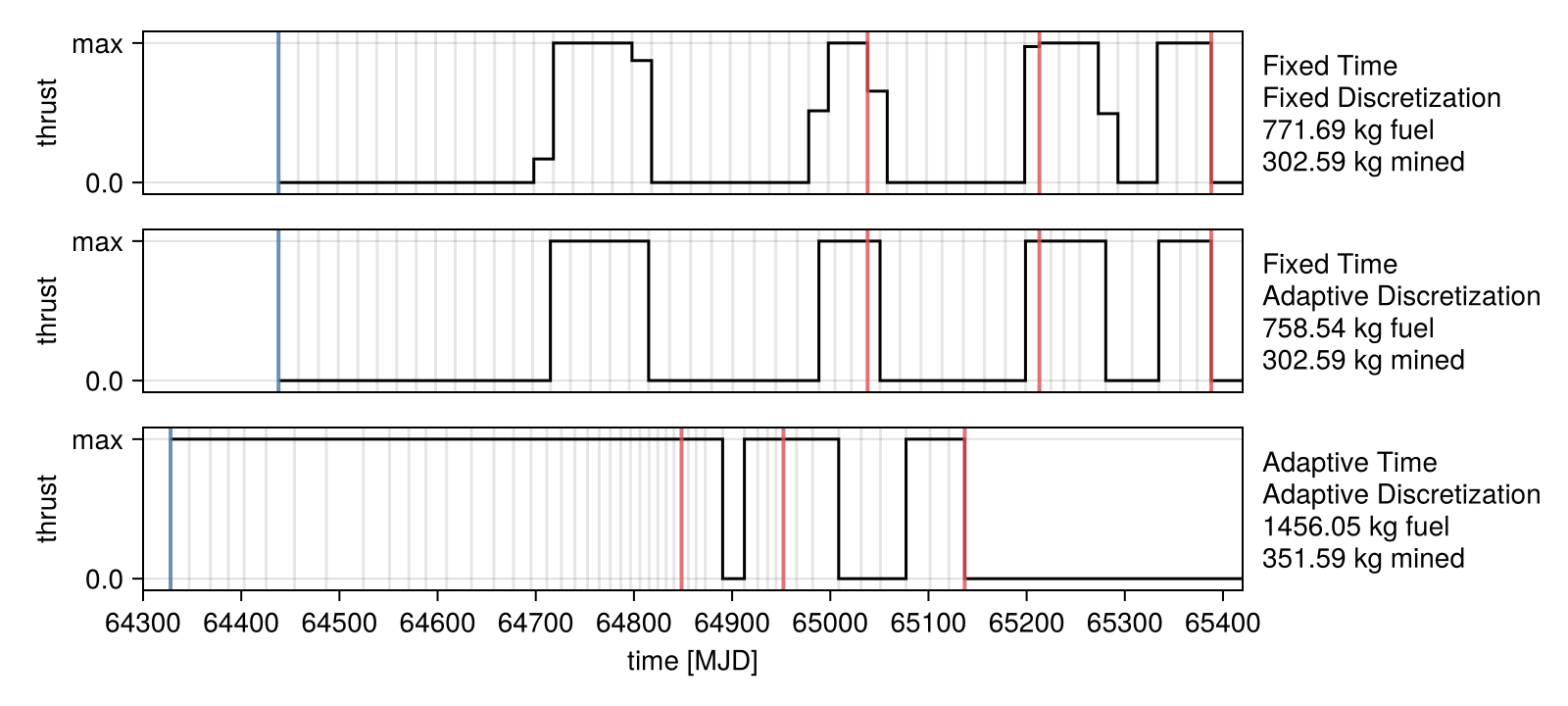}
    \caption{Comparison of discretization and optimal control profile during the deployment phase.}
    \label{discretization_explainer}
\end{figure}

Figure~\ref{discretization_explainer} provides more detail on the behavior of the adaptive-mesh and subsequently the rendezvous time optimization. A close-up of the deployment phase is shown for 3 different optimization methodologies. Firstly, a fixed discretization exhibits a clear control structure, but this is not well defined due to suboptimal time discretization. Next, when the discretization is permitted to adapt, a clearer bang-bang control profile is recovered as is expected in a mass-optimal solution. However, the solution remains suboptimal with respect to our intended objective. Finally, allowing the rendezvous times to change retains the bang-bang control profile, but with rendezvous events occurring significantly earlier. This permits larger amounts of mined mass with the tradeoff of increased fuel use. 

\subsubsection{Multiple trajectory optimization}

The SCP formulation can be adapted to optimize multiple ship trajectories simultaneously, which is essential when ships are interdependent and require joint optimization. Whilst self-cleaning ships in the GTOC 12 problem can be optimized independently because they do not impact each other, mixed ships present an example where simultaneous optimization becomes crucial. This is because different ships may separately handle the miner deployment and mass collection events for an individual asteroid, which couples their time variables. This coupling could influence each individual ship trajectory, so for mixed ship campaigns, all ship trajectories must be optimized simultaneously.

As with the BIP, a separate SCP formulation is created for each ship, and all of the constraints and variables for each ship are integrated into a unified SCP formulation. To manage the additional sets of variables required, those pertaining to each ship are denoted with a superscript index, such as the time variable $t^{\,l}_{i}$. Most of the constraints from the single trajectory SCP (e.g., \eqref{dynamics_linear}, \eqref{control_magnitude}, \eqref{state_violation}) remain applicable but are now applied to each ship's trajectory independently. Particular attention is required for the linearized rendezvous constraints \eqref{rendezvous_dynamics_linear_start} and \eqref{rendezvous_dynamics_linear_end}, as the each rendezvous now depends on the combinatorial ordering of the mixed ships. 

The most significant modification occurs in the mass linkage constraint \eqref{mass_linkage}, which must now account for the interaction between mixed ships. Firstly, the approximation of the mass difference between legs $\Delta m$ now needs to be adjusted to accommodate the mixed ships:
\begin{equation} \label{mass_deltas_mixed}
\begin{aligned} 
    \Delta m^{\,l}_i = \left\{\begin{array}{ll}
            -m_{\text{miner}} & t^{\,l}_{i} \in \text{deployment}\\
            m_{\text{rate}} \left( t^{\,l}_{i} - t_{\text{deployment}(i, l)} \right) & t^{\,l}_{i} \in \text{collection}
    \end{array}\right.
\end{aligned}
\end{equation}
where $t_{\text{deployment}(i, l)}$ represents the time at which the miner deployment occurs at the asteroid visited at the start of leg $i$ for ship $l$. A reference $\Delta \bar{m}^{\,l}_{i}$ is then calculated based on the rendezvous times from the previous iteration of SCP. Consequently, the mass linkage constraint for mixed ships can be expressed as:
\begin{equation} \label{mass_linkage_mixed}
\begin{aligned} 
    (\text{true-mass})\, \forall i, \forall l, i>1:& \, m^{\,l}_{1, i} = m^{\,l}_{\text{end}, i-1} + \Delta m^{\,l}_{i} \\ 
    (\text{logged-mass})\, \forall i, \forall l, i>1:& \, w^{\,l}_{1, i} = \log \left( e^{\bar{w}^{\,l}_{\text{end}, i-1}} + \Delta \bar{m}^{\,l}_{i} \right) + \frac{\left( w^{\,l}_{\text{end}, i-1} -  \bar{w}^{\,l}_{\text{end}, i-1} \right)}{1 + \Delta \bar{m}^{\,l}_{i} / e^{\bar{w}^{\,l}_{\text{end}, i-1}}} 
\end{aligned}
\end{equation}

Similar to the constraint for a single trajectory \eqref{mass_linkage}, the additional terms in the expression for the logged-mass constraint arise from the linearization of $\log \left( e^{w^{\,l}_{\text{end}, i-1}} + \Delta m^{\,l}_{i} \right)$ around a reference $\bar{w}^{\,l}_{\text{end}, i-1}$ from the previous iteration of SCP. 

The objective function should also be adapted to include each ship simultaneously and now aims to maximize the total spread of all times at once:
\begin{align}
    J = \sum_{i, l, t^{\,l}_{i} \in \text{deployment}} t^{\,l}_{i} - \sum_{i, l, t^{\,l}_{i} \in \text{collection}} t^{\,l}_{i} - 10^4 \sum_{i, l} \bm{\delta}_{i, \text{dep}}^{+, \,l} - 10^4 \sum_{i, l} \bm{\delta}_{i, \text{arr}}^{+, \,l} - 5\times 10^3 \sum_l \gamma^{\,l}
\end{align}

By maximizing the spread of all the times, this objective allows for the individual objectives of some ship trajectories to worsen, provided that the overall time spread across all ships is increased. For instance, a particular trajectory might underperform due to a significant mass pickup from an asteroid, which in turn increases fuel consumption for subsequent legs. By altering the deployment timing of the miner on that asteroid in another trajectory (thus decreasing that particular ship's objective), it may lead to more efficient subsequent transfers and allow for a greater increase in overall objective when compared to the individual decrease incurred.

Finally, the convergence of this SCP formulation is checked in a manner similar to the optimization of a single trajectory but is applied collectively across all trajectories. It is worth noting that it is possible for a single trajectory to meet the convergence criterion initially, only to fail to do so in subsequent iterations due to changes in the mass collected at collection events caused by changes in deployment time across the mixed ship campaign. Therefore, the optimization is considered converged only once all trajectories simultaneously meet the convergence criterion.

\subsection{Campaign assembly}

The final step of the process to create a complete campaign of ship trajectories for the GTOC 12 problem is the selection process. After running the nested-loop process multiple times with different inputs, a large pool of possible ship trajectories is generated. The goal is to select the best combination of these ships whilst ensuring that there are no conflicts where multiple ships deploy or collect from the same asteroids.

This selection process is achieved using a separate BIP formulation, which supports the optimization of self-cleaning, mixed, and hybrid campaigns containing both types of ships. Firstly, a binary variable $s_i \in [0, 1]$ is introduced to represent whether ship $i$ is selected. Several constraints are then enforced to ensure that only non-conflicting ships are chosen for the campaign. These constraints are detailed below.

\textbf{(whole mixed campaign selection)} When mixed ships are included in the campaign, each set must be included all at once. This is specified by first arranging each ship index $i$ into $j$ groups to form $g_j$, such that each self-cleaning ship has its own group while mixed ships are within the same group. Then, the following constraint ensures all of the ship indices within a group assume the same value:
\begin{align} \label{mixed_ships}
    \forall j,  \forall i\in g_j,  \forall k\in g_j: s_i = s_k
\end{align}

\textbf{(maximum one deployment/collection per asteroid)} Next, consider each possible target asteroid $a_k$, and define a function $f(a_k, i)$ that returns the number of times asteroid $a_k$ is visited by ship $i$. The following constraint then ensures that each asteroid is visited no more than twice (once for deployment and once for collection):
\begin{align} \label{single_deploy_collect}
    \forall k: \sum_{i} f(a_k, s_i) \leq 2
\end{align}

This leaves the possibility of an asteroid being visited only once (for either deployment or collection), but when combined with \eqref{mixed_ships}, this issue is avoided, assuming mixed campaigns correctly deploy and collect from all assigned asteroids.

\textbf{(campaign ship number limit)} Finally, the total number of ships in the campaign is limited based on problem requirements. Here, the total number of ships in the campaign $N$ is left as an integer parameter to the BIP, allowing for sequential BIP re-runs until the mass limitation requirement is no longer satisfied:
\begin{align} \label{ship_number}
    \sum_{i} s_i \leq N
\end{align}

The objective is to maximize the total returned mass by the selected ships, where $m_i$ represents the total mined mass returned by ship $i$:
\begin{align} \label{selection_objective}
    J = \sum_{i} m_{i} s_{i}
\end{align}

Alternatively, this objective can be modified to support the penalized mass criterion from GTOC 12 \cite{gtoc12ProblemDescription}, which was used as the objective in the competition to rank the solutions. In either case, the complete BIP program for ship selection is:
\begin{mini}[2]
    {s_{i}}{\eqref{selection_objective}}{}{}
    \addConstraint{\eqref{mixed_ships}}{}{\quad\text{(whole mixed campaign selection)}}
    \addConstraint{\eqref{single_deploy_collect}}{}{\quad\text{(maximum one deployment/collection per asteroid)}}
    \addConstraint{\eqref{ship_number}}{}{\quad\text{(campaign ship number limit)}}
    \addConstraint{s_{i}}{\in \{0, 1\}}{\quad\text{(binary variables)}}
    \label{bip_ship_selection}
\end{mini}

\section{\label{Results}Results}
In this section, the process to construct high-quality self-cleaning and mixed strategies for GTOC 12 is outlined and demonstrated using the proposed methodology. Firstly, the construction of a high-quality self-cleaning solution is demonstrated from a restricted subset of asteroid targets. This is followed by a similar process for constructing a mixed set of solutions. Finally, 37-ship self-cleaning and 39-ship mixed campaigns are presented, which have been primarily constructed through this methodology. These campaigns currently represent the best-known solutions to the GTOC 12 problem.

All algorithms were written and implemented using version 1.11 of the Julia programming language and executed using an Intel Core i7-12700 processor with 32 GB of RAM.

\subsection{\label{reoptimization_solutions}Rendezvous time optimization of top 10 solutions}

\begin{figure}
    \centering
    \includegraphics[width = \textwidth]{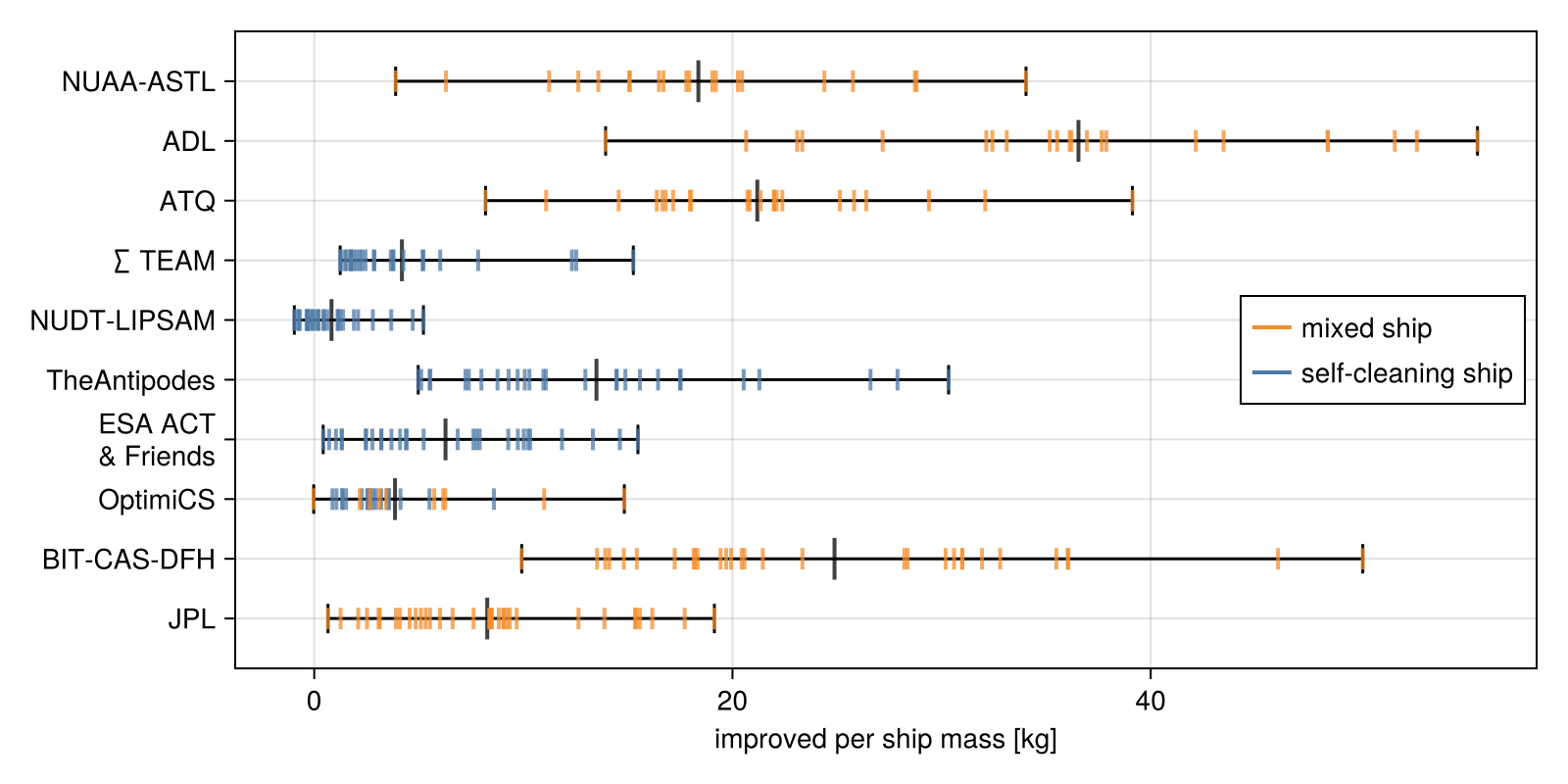}
    \caption{Application of adaptive-time SCP to the top 10 teams' solutions submitted within the GTOC 12 timeframe.}
    \label{top10_gtoc_reoptimized}
\end{figure}

To evaluate the benefits of incorporating rendezvous time optimization directly into the trajectory optimization, the solutions from the top 10 submissions to the GTOC 12 problem are reoptimized. The inputs to the re-optimization process include the rendezvous sequence and prior rendezvous times, which served as initial guesses for the adaptive-mesh SCP implementation. The output consisted of valid optimal control profiles for each ship.

The results of this reoptimization are illustrated in Figure~\ref{top10_gtoc_reoptimized}. Notably, it is apparent that almost all of the submitted solutions to GTOC 12 exhibit a considerable degree of suboptimality regarding the selection of rendezvous times. In certain instances, especially within teams employing mixed campaigns, the increased average mined mass could have facilitated the inclusion of several additional ships within the team's solutions.

This re-optimization presented an excellent opportunity to assess the computational performance of the adaptive-mesh SCP, given the large number of solutions to optimize and the varying methodologies employed by different teams. Our tests indicated that the adaptive-mesh SCP efficiently handles all teams' solutions within a reasonable timeframe. Most re-optimizations completed in about 10 minutes starting from an initial time discretization guess of 20 days. 

As an example, consider the solution submitted by JPL, which consists of 35 mixed ships, each performing an average of 18 rendezvous events corresponding to 20 low-thrust legs. For each SCP iteration, the convex subproblem involved 136,385 variables and 180,421 constraints. Then, the total runtime was 13.18 minutes, yielding an average re-optimization time of 22.60 seconds per ship. 

\subsection{\label{single_ship_results}Self-cleaning ship optimization}

\begin{figure}[t]
    \centering
    \includegraphics[width = 0.62\textwidth]{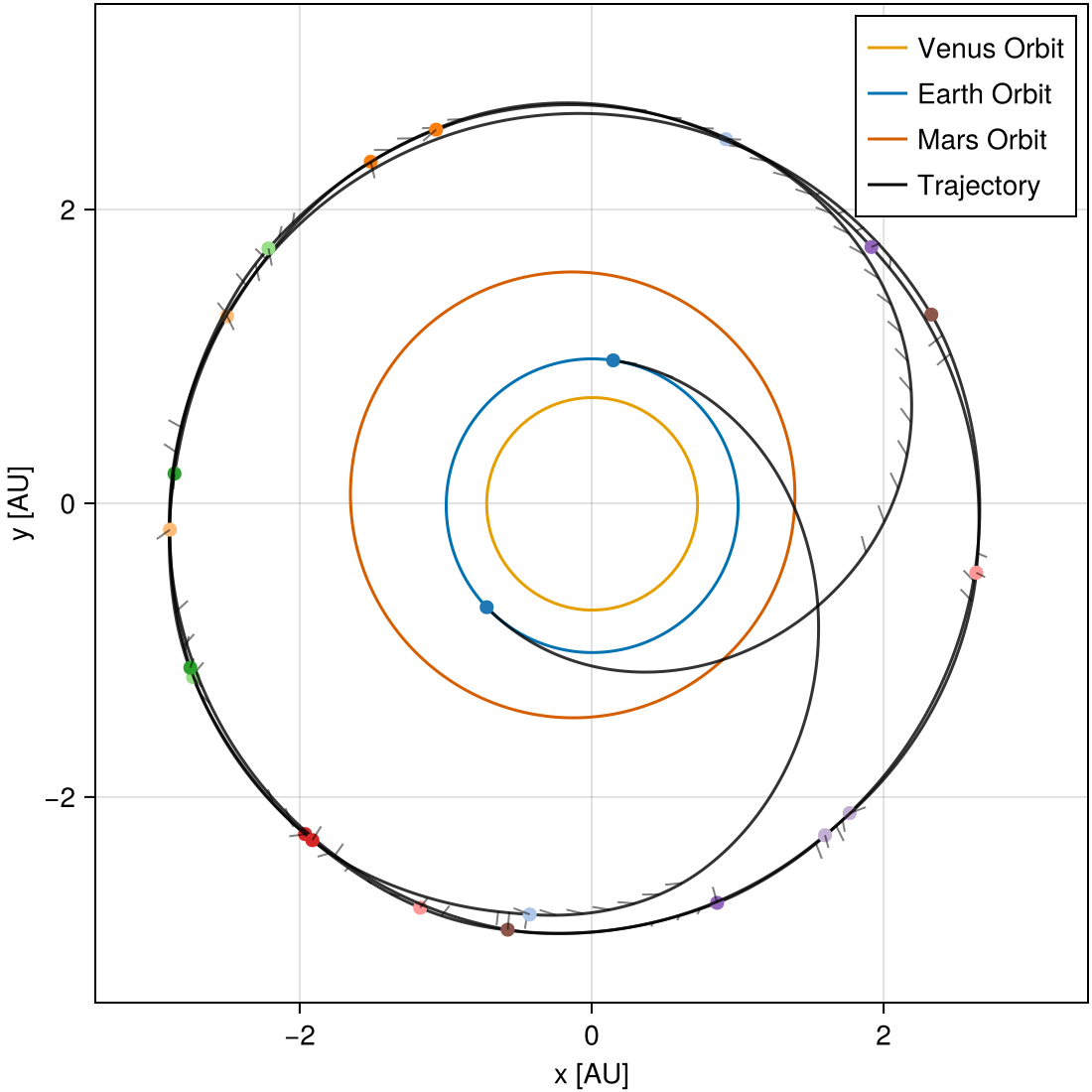}
    \caption{Best self-cleaning ship trajectory found, returning 781 kg of mined mass.}
    \label{individual_trajectory}
\end{figure}

\begin{table}
    \centering
    \caption{Initial rendezvous time guess and optimized solution for the best self-cleaning ship found.}
    \begin{tabular}{lrrrrr}\toprule
        & \multicolumn{2}{l}{\textbf{Initial Guess}} & \multicolumn{3}{l}{\textbf{Optimized Solution}} \\
        \cmidrule(l){2-3} \cmidrule(l){4-6} 
    
        \textbf{Event} & \textbf{Time} $\left[\textbf{MJD}\right]$ & $\mathbf{\Delta t} \left[\textbf{days}\right]$ & \textbf{ID} & \textbf{Time} $\left[\textbf{MJD}\right]$ & $\mathbf{\Delta t} \left[\textbf{days}\right]$\\
        \midrule
        Earth Departure & $64438.00$ & - & - & $64452.37$ & -\\
        
        Miner Deployment 1 & $65038.00$ & $600.00$ & 15184 & $64962.05$ & $509.68$\\
        Miner Deployment 2 & $65183.00$ & $145.00$ & 3241 & $65217.95$ & $255.90$\\ 
        Miner Deployment 3 & $65328.00$ & $145.00$ & 32088 & $65358.55$ & $140.60$\\
        Miner Deployment 4 & $65473.00$ & $145.00$ & 23987 & $65469.87$ & $111.32$\\
        Miner Deployment 5 & $65618.00$ & $145.00$ & 23056 & $65610.32$ & $140.45$\\
        Miner Deployment 6 & $65763.00$ & $145.00$ & 46751 & $65745.12$ & $134.80$\\
        Miner Deployment 7 & $66053.00$ & $290.00$ & 2032 & $65845.86$ & $100.74$\\
        Miner Deployment 8 & $66343.00$ & $290.00$ & 19702 & $66056.93$ & $211.07$\\
        Miner Deployment 9 & $66633.00$ & $290.00$ & 46418 & $66163.66$ & $106.73$\\
        Miner Deployment 10 & $66923.00$ & $290.00$ & 53592 & $66501.42$ & $337.76$\\
        
        Mass Collection 1 & $67347.00$ & $424.00$ & 53592 & $67601.67$ & $1100.25$\\
        Mass Collection 2 & $67637.00$ & $290.00$ & 46418 & $67833.63$ & $231.96$\\ 
        Mass Collection 3 & $67927.00$ & $290.00$ & 2032 & $68040.42$ & $206.79$\\
        Mass Collection 4 & $68217.00$ & $290.00$ & 19702 & $68265.09$ & $224.67$\\
        Mass Collection 5 & $68507.00$ & $290.00$ & 3241 & $68578.90$ & $313.81$\\
        Mass Collection 6 & $68652.00$ & $145.00$ & 23056 & $68714.11$ & $135.21$\\
        Mass Collection 7 & $68797.00$ & $145.00$ & 32088 & $68910.77$ & $196.66$\\
        Mass Collection 8 & $68942.00$ & $145.00$ & 23987 & $69006.37$ & $95.60$\\
        Mass Collection 9 & $69087.00$ & $145.00$ & 46751 & $69161.79$ & $155.42$\\
        Mass Collection 10 & $69232.00$ & $145.00$ & 15184 & $69324.37$ & $162.58$\\
        
        Earth Arrival & $69782.00$ & $550.00$ & - & $69787.32$ & $462.95$\\
        \bottomrule
    \end{tabular}
    \label{best_single_details}
\end{table}

The construction process of the best self-cleaning ship identified in our analysis, which returns 781 kg of mined mass, is demonstrated. This solution is derived from a restricted subset of 20 asteroids and includes 10 deployment and collection events. In our initial analysis of the same subset, we found that solutions with 8 to 9 deployment/collection events performed well, yielding approximately 730 kg and 760 kg, respectively. Thus, the addition of a 10th deployment and collection event led to the best self-cleaning solution found. This is visualized in Figure~\ref{individual_trajectory}, and further details can be found in Table~\ref{best_single_details}. 

The choice of initial rendezvous time schedule was critical to ensuring that the nested-loop algorithm converged to a solution exhibiting characteristics similar to those found in other high-quality solutions. For the initial guess of the rendezvous times, we constructed the schedule with an initial spacing of 145 days for the first and last five transfers, while all remaining transfers were spaced at 290 days. This approach ensures that several of the first and last transfers are conducted quickly, allowing more time for material to be mined on average. Detailed information on the input rendezvous time schedule can be found in Table~\ref{best_single_details}.

\begin{figure}[h!]
    \centering
    \includegraphics[width = \textwidth]{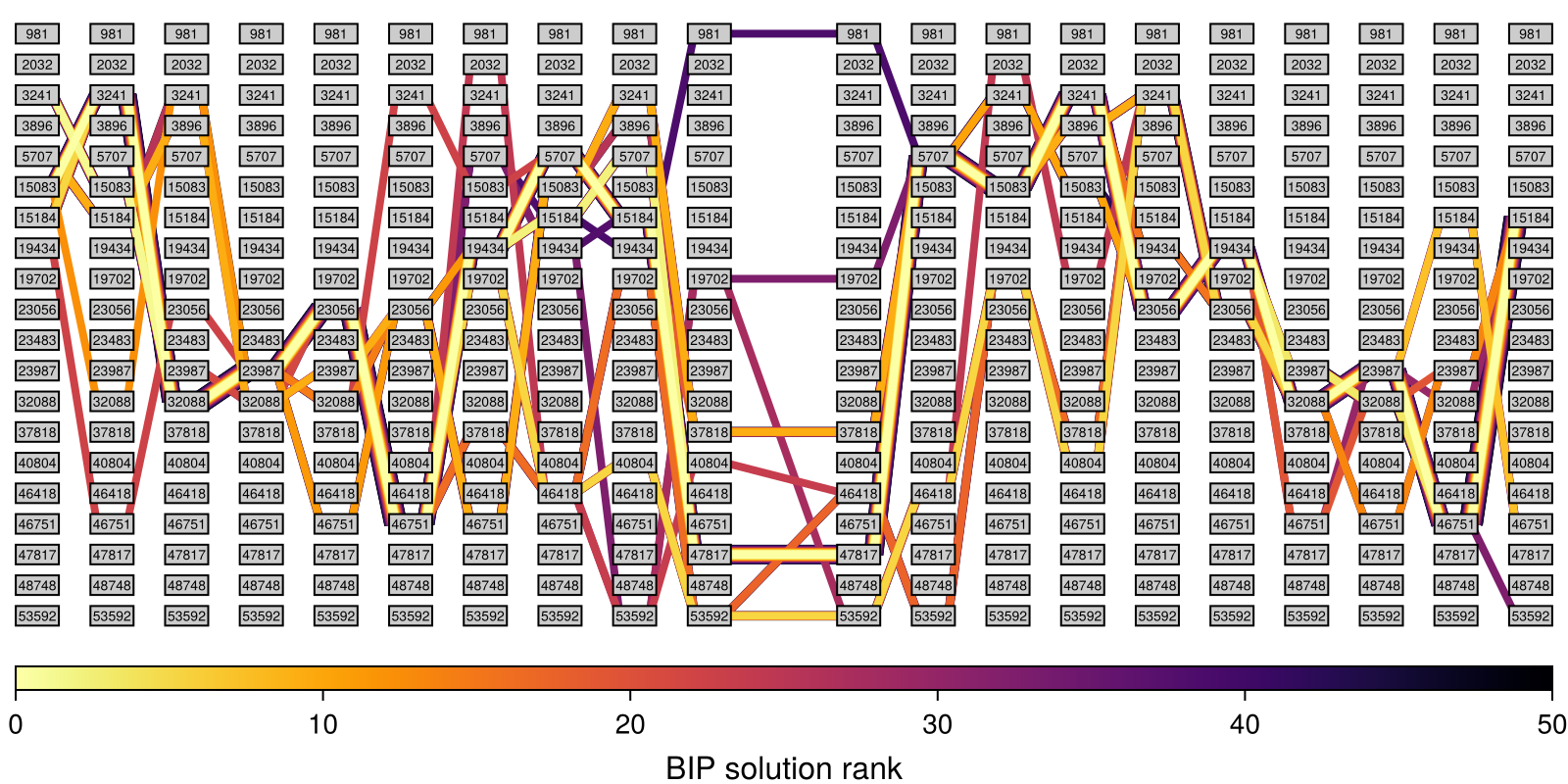}
    \caption{BIP solution structure on the first iteration solving for the self-cleaning ship found.}
    \label{single_ship_network}
\end{figure}

Next, we demonstrate the intermediate details of the algorithm. The BIP solution structure for a single ship is visualized in Figure~\ref{single_ship_network}, with only active arcs shown for clarity. Additionally, the top 50 solutions to the BIP are displayed and colored according to their ranking. It is evident that many of the top solutions exhibit significant overlap, and many solutions remain at the same asteroid between the deployment and collection phases.

In comparison to previous example cases (e.g., Figure~\ref{bip_solutions}), this structure is denser, containing a great number more variables. The pruning cutoff was 6 km/s, resulting in the pruning step leaving 4,780 variables out of a maximum possible of 7,600. Despite this significantly increased complexity, this formulation was still able to be solved within a reasonable timeframe, achieving proven optimality in 22.46 seconds to find the 50 best combinatorial orderings.

\begin{figure}
    \centering
    \includegraphics[width = \textwidth]{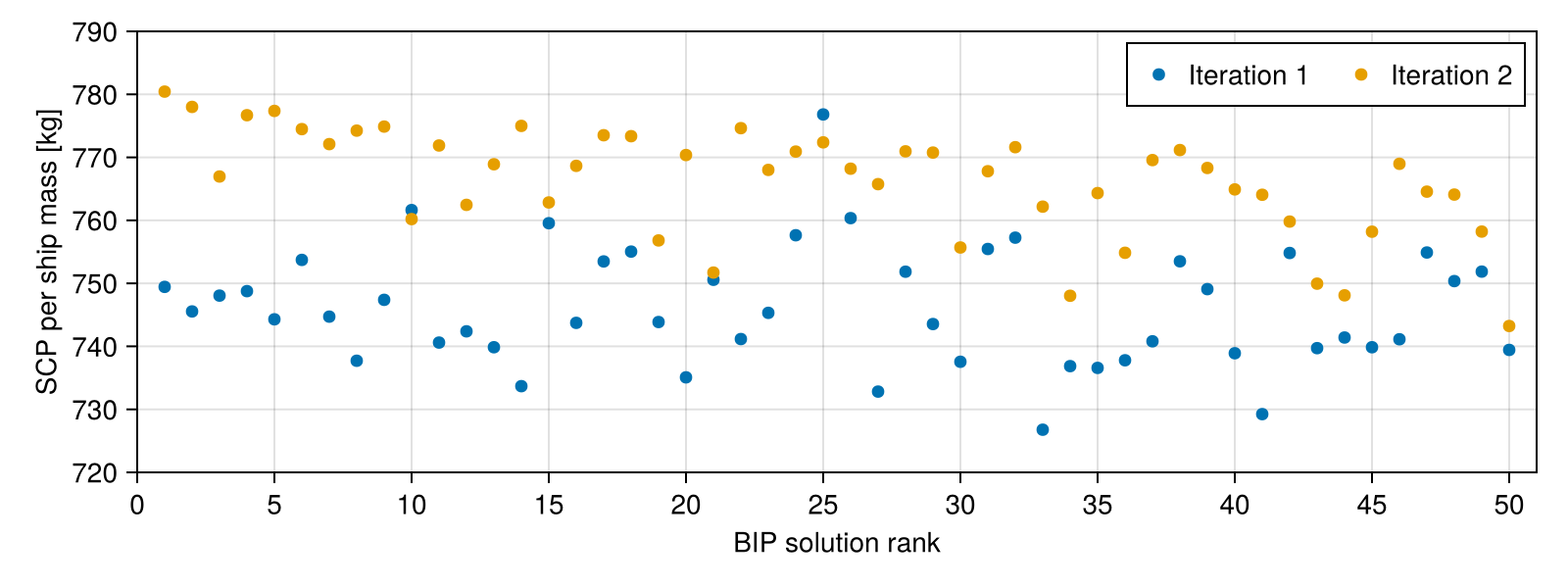}
    \caption{Progress of the nested-loop algorithm when solving the individual ship problem.}
    \label{nested_loop_progress}
\end{figure}

Next, the adaptive-mesh SCP is applied to all 50 combinatorial orderings generated by the BIP step, the results of which can be seen in Iteration 1 of Figure~\ref{nested_loop_progress}. This process took 15.22 minutes, averaging 18.26 seconds per ship. Remarkably, all of the solutions found by the first BIP step proved feasible and are of high quality, with an average return of 746.04 kg and the best solution returning 776.81 kg. Notably, the best solution found in the first iteration was only the 25th-ranked solution from the BIP, underscoring the significant impact of the approximations made to the BIP objective - such as transfer costs determined via Lambert transfers.

The nested-loop process then repeats the BIP and SCP steps, as shown in Iteration 2 of Figure~\ref{nested_loop_progress}. In this second iteration, the BIP rankings better reflect the true solution quality; notably, the top-ranked solution of the BIP optimized through the SCP yields the solution detailed in Table~\ref{best_single_details} which returns 781 kg. Further iterations of the nested-loop process showed very similar behavior to Iteration 2, providing no further improvements. When testing the nested-loop process on different initial guesses of rendezvous time schedule, similar results were obtained. However, these tended to take more iterations of the nested-loop process to converge, or sometimes tended to converge to suboptimal local solutions that returned approximately 750 kg. 

\subsection{\label{mixed_ship_results}Mixed ship optimization}

\begin{figure}[t]
    \centering
    \includegraphics[width = \textwidth]{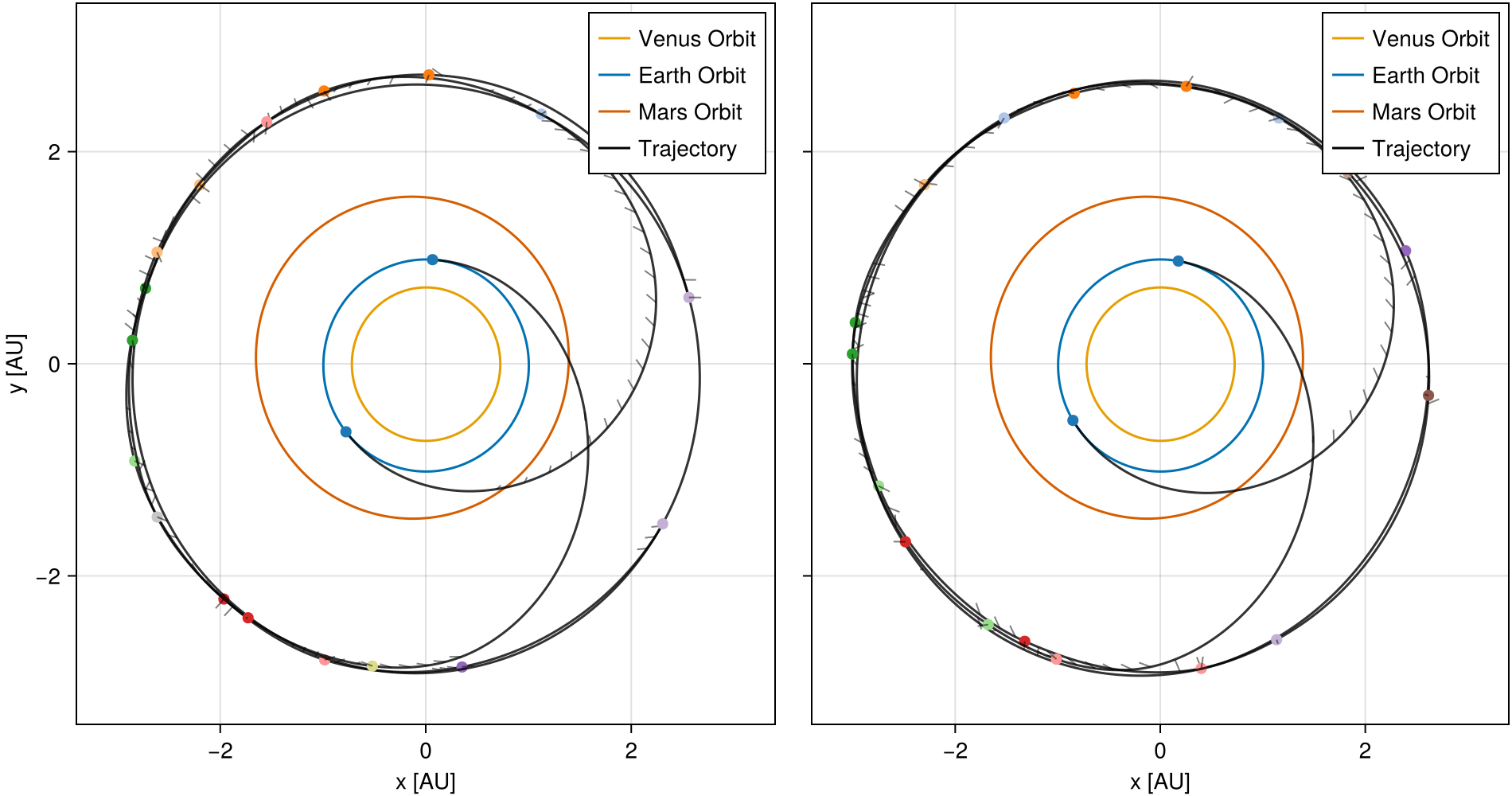}
    \caption{Pair of mixed ship trajectories, the first ship (left) returns 681.75 kg of mined mass and the second ship (right) returns 777.33 kg of mined mass.}
    \label{mixed_trajectory}
\end{figure}

\begin{table}
    \centering
    \caption{Optimized solution for the pair of mixed ships example.}
    \begin{tabular}{lrrrrrr}\toprule
        & \multicolumn{3}{l}{\textbf{Ship 1}} & \multicolumn{3}{l}{\textbf{Ship 2}} \\
        \cmidrule(l){2-4} \cmidrule(l){5-7} 
    
        \textbf{Event} & \textbf{ID} & \textbf{Time} $\left[\textbf{MJD}\right]$ & $\mathbf{\Delta t} \left[\textbf{days}\right]$ & \textbf{ID} & \textbf{Time} $\left[\textbf{MJD}\right]$ & $\mathbf{\Delta t} \left[\textbf{days}\right]$\\
        \midrule
        Earth Departure & - & $64451.48$ & - & - & $64443.76$ & -\\
        
        Miner Deployment 1 & 15083 & $64989.72$ & $538.24$ & 19702 & $64977.04$ & $533.28$\\
        Miner Deployment 2 & 3241 & $65204.69$ & $214.97$ & 46418 & $65132.11$ & $155.07$\\ 
        Miner Deployment 3 & 32088 & $65345.32$ & $140.63$ & 47817 & $65272.00$ & $139.89$\\
        Miner Deployment 4 & 23987 & $65456.37$ & $111.05$ & 9260 & $65437.51$ & $165.51$\\
        Miner Deployment 5 & 23056 & $65600.10$ & $143.73$ & 981 & $65592.68$ & $155.17$\\
        Miner Deployment 6 & 46751 & $65736.37$ & $136.27$ & 17983 & $65801.44$ & $208.76$\\
        Miner Deployment 7 & 2032 & $65844.52$ & $108.15$ & 19434 & $65965.22$ & $163.78$\\
        Miner Deployment 8 & 39557 & $66097.98$ & $253.46$ & 5707 & $66282.35$ & $317.13$\\
        Miner Deployment 9 & 48748 & $66279.64$ & $181.66$ & - & - & - \\
        
        Mass Collection 1 & 48748 & $68117.36$ & $1837.72$ & 5707 & $67732.70$ & $1450.35$\\
        Mass Collection 2 & 2032 & $68365.46$ & $248.10$ & 15083 & $68087.79$ & $355.09$\\ 
        Mass Collection 3 & 46418 & $68571.65$ & $206.19$ & 3241 & $68425.12$ & $337.33$\\
        Mass Collection 4 & 47817 & $68685.79$ & $114.14$ & 23056 & $68614.45$ & $189.33$\\
        Mass Collection 5 & 9260 & $68897.51$ & $211.72$ & 32088 & $68794.10$ & $179.65$\\
        Mass Collection 6 & 17983 & $69095.28$ & $197.77$ & 23987 & $68885.64$ & $91.54$\\
        Mass Collection 7 & 981 & $69207.52$ & $112.24$ & 46751 & $69023.47$ & $137.83$\\
        Mass Collection 8 & 19434 & $69285.50$ & $77.98$ & 19702 & $69191.97$ & $168.50$\\
        Mass Collection 9 & - & - & - & 39557 & $69317.35$ & $125.38$\\
        
        Earth Arrival & - & $69785.77$ & $500.27$ & - & $69791.50$ & $474.15$\\
        \bottomrule
    \end{tabular}
    \label{best_mixed_details}
\end{table}

The creation of mixed ships is demonstrated using a subset similar to that of the self-cleaning example but expanded to include 30 asteroids. In this scenario, a set of two mixed ships is desired: one ship is configured to have 8 deployments and 9 collections, while the other has 9 deployments and 8 collections.

The initial rendezvous schedule mirrors that of the self-cleaning example, though intermediate transfers are removed in accordance with the specified number of deployments and collections (details can be found in Table~\ref{best_single_details}). The pruning cutoff remains at 6 km/s, resulting in a pruning step that now leaves 15,122 variables out of a maximum possible of 28,800. This BIP solution is significantly more computationally demanding than its self-cleaning counterpart due to the increased number of variables and the inclusion of the second ship into the problem structure. With a maximum allowable time of 20 minutes to solve, the BIP achieves an optimality gap of 5.91\%. Although it is no longer provable that the obtained solutions are the best ranked, the low optimality gap indicates that the solutions are still of high quality.

Subsequently, the SCP is run on all the mixed solutions obtained from the BIP step, taking a total of 21.95 minutes, averaging 26.34 seconds per set of mixed ships. In later iterations of the nested-loop process, the optimal solution to the BIP is obtained much more quickly (for instance, within 13 minutes in the second iteration) rather than just achieving an optimality gap. This improvement is likely because the optimal solution becomes more clearly defined once the rendezvous times are optimized by the SCP step.

Upon convergence of the nested-loop process, a solution returning 729.54 kg of mined mass on average per ship is achieved. This solution is presented in Figure~\ref{mixed_trajectory}, with specific details available in Table~\ref{best_mixed_details}. Notably, the solution demonstrates significant mixing between the two ships; for instance, out of the 9 deployments made by the first ship, the second ship is responsible for collecting from 7 of those deployments.

\subsection{\label{solution_assembly}Creation of new GTOC 12 solutions}

A large pool of potential solutions was created by separately running the nested-loop process on various subsets of asteroids, utilizing a range of different starting rendezvous time guesses. This pool was subsequently combined with the pool of reoptimized solutions from the top 10 team's submissions, resulting in a comprehensive solution pool comprising a total of 589 ship trajectories (358 self-cleaning and 231 mixed).

\begin{figure}[t]
    \centering
    \includegraphics[width = \textwidth]{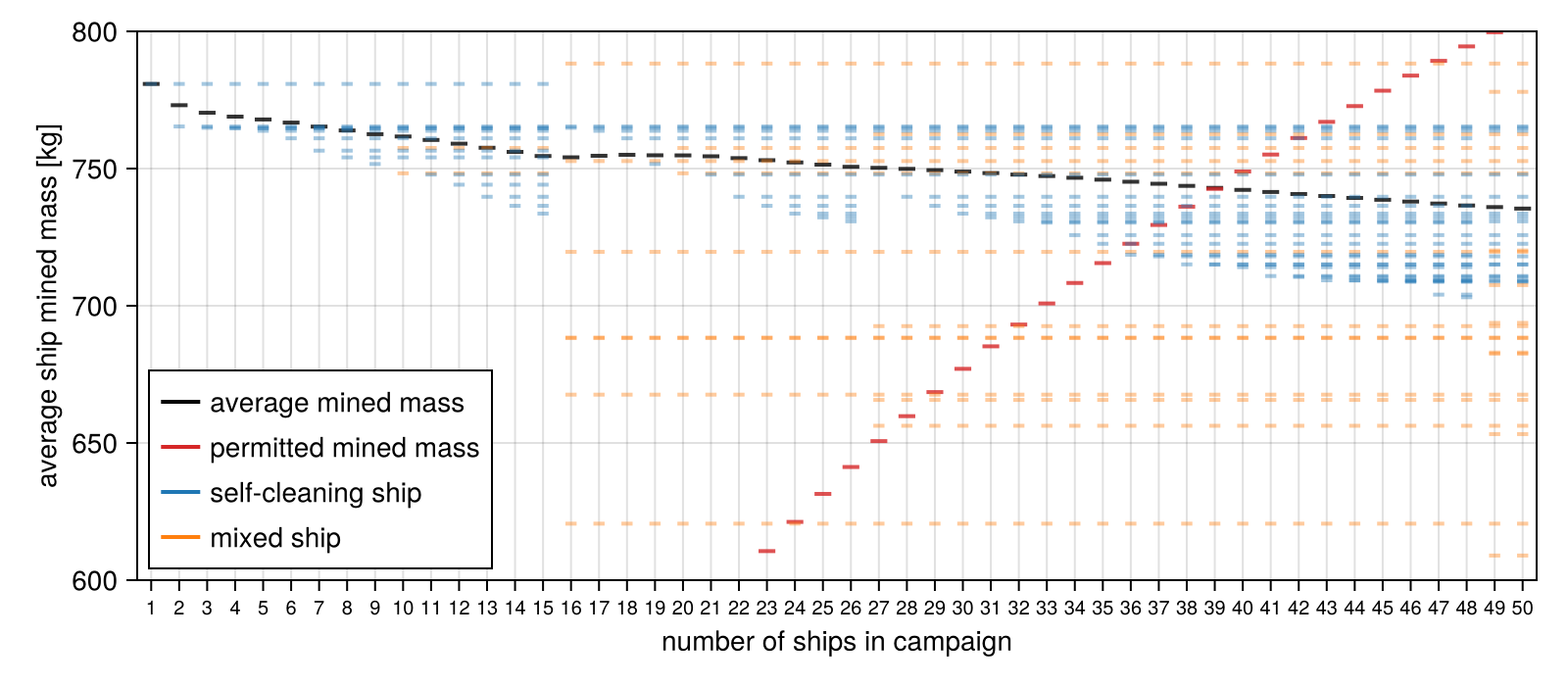}
    \caption{Optimal ship campaign selections for the solution pool from size 1 to 50.}
    \label{ship_selection}
\end{figure}

The results of the campaign assembly process applied to this pool are visualized in Figure~\ref{ship_selection}, showcasing optimal ship campaigns ranging in size from 1 to 50 ships. Using the Gurobi solver, the calculations for all campaigns took a total of just 0.95 seconds, clearly demonstrating the good computational performance of the campaign assembly process. This efficiency allowed for rapid exploration of various campaign configurations, for example, those only including self-cleaning ships.

\begin{figure}[t]
    \centering
    \includegraphics[width = \textwidth]{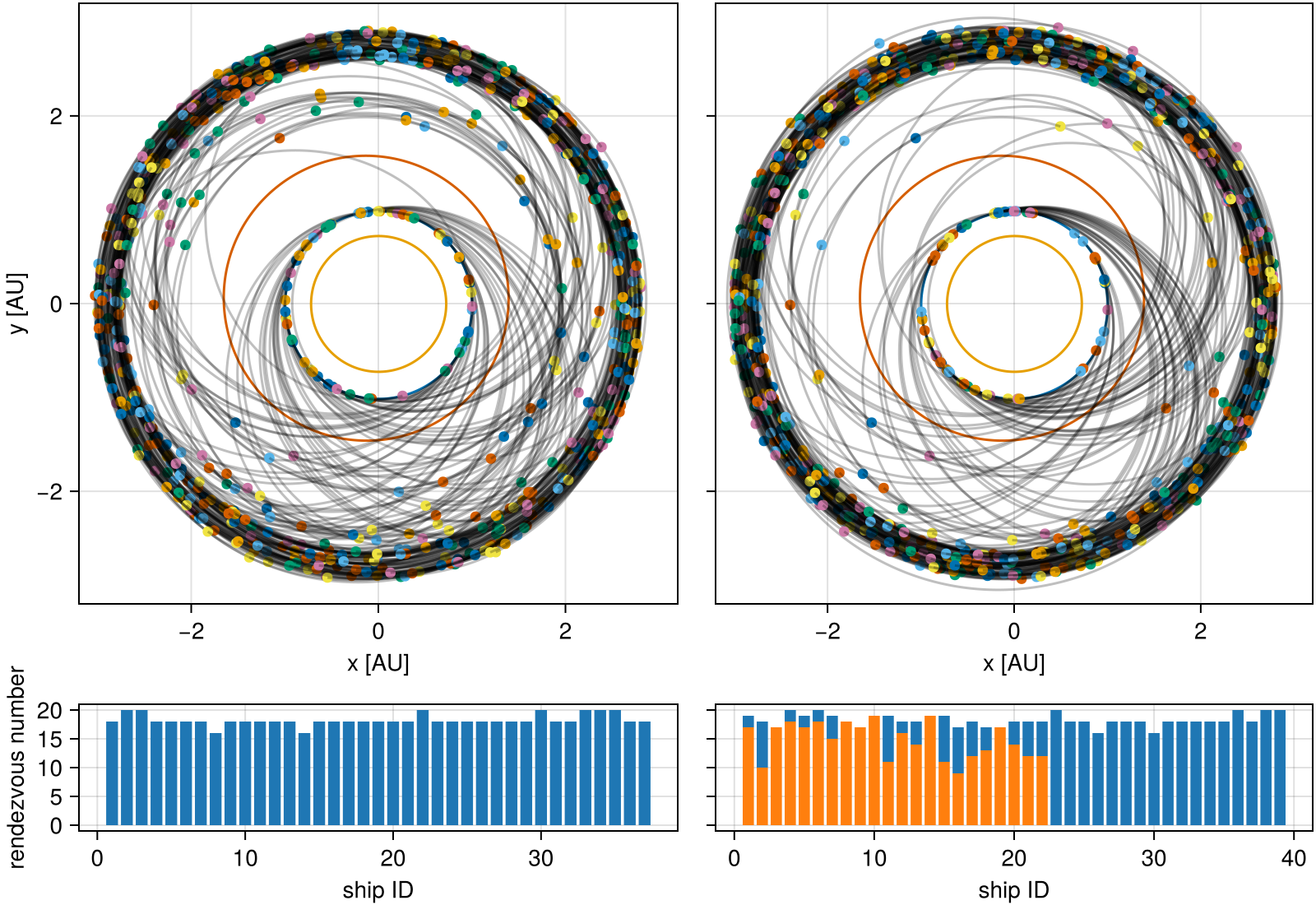}
    \caption{Best combined self-cleaning (left) and mixed (right) solutions. The best self-cleaning solution is a campaign of 37 ships, and the best with mixed ships is a campaign of 39 ships. In the bottom panels, self-cleaning rendezvous are shown in blue and mixed in orange.}
    \label{combined_both}
\end{figure}

Using this solution pool and the assembly optimization process, new best solutions have been obtained for the GTOC 12 problem. Figure~\ref{combined_both} illustrates these solutions: the first is a 37-ship solution composed entirely of self-cleaning ships developed using this methodology, achieving an average mined mass of 730.95 kg per ship. The second solution is a 39-ship configuration consisting of 17 self-cleaning ships created through this methodology and 20 mixed ships, which were primarily modified from the JPL solution. This 39-ship campaign yields an average mined mass of 742.95 kg per ship. The verifiable files for these solutions are available at \url{https://sophia.estec.esa.int/gtoc_portal/?page_id=1261}.

To add an additional ship to the self-cleaning campaign, the average mined mass would need to increase to 736.11 kg per ship, while the current average stands at 730.07 kg with 38 ships. This means an additional 229.52 kg of mined mass would be required across the entire 38-ship campaign to make this addition feasible.

In the case of the mixed campaign, an increase in average mined mass to 748.93 kg per ship would be necessary, whereas the current average is 742.23 kg with 40 ships. Thus, a total of 268.00 kg of extra mined mass would be needed across the entire 40-ship campaign to permit the addition of another ship.

\section{\label{sec:Conclusion}Conclusions}

This work presented a nested-loop optimization framework, combining Binary Integer Programming (BIP) and adaptive-mesh Sequential Convex Programming (SCP), which has been demonstrated to effectively address certain types of multi-target spacecraft trajectory problems. Applying this methodology to the 12th Global Trajectory Optimization Competition (GTOC 12) problem led to new best-known solutions, including 37-ship and 39-ship campaigns with mined mass per ship exceeding the performance of previously submitted solutions, thereby validating the effectiveness of the approach. Furthermore, the relative simplicity of the framework, particularly in its nested-loop structure, makes it both accessible and straightforward to implement.

The application of adaptive-mesh SCP to previously submitted solutions for GTOC 12 identified significant suboptimalities in rendezvous timing, resulting in substantial performance gains. These improvements not only highlight the critical role of rendezvous time optimization to multi-target problems, but also help to enhance the effectiveness of the BIP process within the nested-loop framework. By refining the time schedule input to the BIP, the adaptive-mesh SCP enables the BIP to generate more accurate and efficient target selection, explaining how the two components mutually reinforce each other to enable superior results compared to existing solutions.

While the need for problem-specific tuning to the initial rendezvous time schedule remains a limitation, this does not detract from the broader applicability of the framework, wherein a suitable approximation to rendezvous times is likely to already be known. The integration of more accurate transfer cost estimates into this framework would further reinforce its potential by improving the accuracy of the BIP approximations, and would be the object of future study.

% \section*{Appendix}

% An Appendix, if needed, appears \textbf{before} research funding information and other acknowledgments.

\section*{Acknowledgments}

The authors would like to thank all members of the GTOC 12 team, `TheAntipodes'. Harry Holt's work was supported by the Warwick \& Judy Smith Engineering Endowment Fund, the University of Auckland Foundation, and the Faculty of Engineering. Since June 2024, Harry Holt has been working as a Research Fellow at ESA's Advanced Concepts Team.

\bibliographystyle{elsarticle-num} 
\bibliography{library.bib} 

\end{document}

%% file: figures/gtoc_explainer.tex
\tikzset{every picture/.style={line width=0.75pt}} %set default line width to 0.75pt        

\begin{tikzpicture}[x=0.75pt,y=0.75pt,yscale=-1,xscale=1]
%uncomment if require: \path (0,317); %set diagram left start at 0, and has height of 317

%Straight Lines [id:da1474818457596283] 
\draw [color={rgb, 255:red, 225; green, 87; blue, 89 }  ,draw opacity=1 ][line width=1.5]    (47.67,235) -- (116.71,224.6) ;
\draw [shift={(120.67,224)}, rotate = 171.43] [fill={rgb, 255:red, 225; green, 87; blue, 89 }  ,fill opacity=1 ][line width=0.08]  [draw opacity=0] (11.61,-5.58) -- (0,0) -- (11.61,5.58) -- cycle    ;
%Straight Lines [id:da014623704836826734] 
\draw [color={rgb, 255:red, 225; green, 87; blue, 89 }  ,draw opacity=1 ][line width=1.5]    (57.33,80.4) -- (110,45.54) ;
\draw [shift={(113.33,43.33)}, rotate = 146.5] [fill={rgb, 255:red, 225; green, 87; blue, 89 }  ,fill opacity=1 ][line width=0.08]  [draw opacity=0] (11.61,-5.58) -- (0,0) -- (11.61,5.58) -- cycle    ;
%Shape: Rectangle [id:dp3844268093670027] 
\draw  [color={rgb, 255:red, 186; green, 176; blue, 172 }  ,draw opacity=1 ][dash pattern={on 3.75pt off 3pt on 7.5pt off 1.5pt}][line width=1.5]  (150,180) -- (540,180) -- (540,280) -- (150,280) -- cycle ;
%Shape: Rectangle [id:dp33515473750794045] 
\draw  [color={rgb, 255:red, 186; green, 176; blue, 172 }  ,draw opacity=1 ][dash pattern={on 3.75pt off 3pt on 7.5pt off 1.5pt}][line width=1.5]  (150.49,30) -- (540.49,30) -- (540.49,130) -- (150.49,130) -- cycle ;
%Curve Lines [id:da8237061445779685] 
\draw [line width=1.5]    (504.53,58.77) .. controls (631.53,16.77) and (626.53,315.11) .. (502.53,252.11) ;
\draw [shift={(597.49,168.84)}, rotate = 271.7] [fill={rgb, 255:red, 0; green, 0; blue, 0 }  ][line width=0.08]  [draw opacity=0] (11.61,-5.58) -- (0,0) -- (11.61,5.58) -- cycle    ;
%Curve Lines [id:da24992335280979772] 
\draw [line width=1.5]    (205.87,251.44) .. controls (159.87,276.44) and (113.2,223.85) .. (53.2,232.11) ;
\draw [shift={(123.45,243.92)}, rotate = 20.03] [fill={rgb, 255:red, 0; green, 0; blue, 0 }  ][line width=0.08]  [draw opacity=0] (11.61,-5.58) -- (0,0) -- (11.61,5.58) -- cycle    ;
%Curve Lines [id:da5207453260865278] 
\draw [line width=1.5]    (397,225.74) .. controls (291,260.47) and (295,198.2) .. (230,240) ;
\draw [shift={(305.06,232.2)}, rotate = 14.1] [fill={rgb, 255:red, 0; green, 0; blue, 0 }  ][line width=0.08]  [draw opacity=0] (11.61,-5.58) -- (0,0) -- (11.61,5.58) -- cycle    ;
%Curve Lines [id:da23897869525605864] 
\draw [line width=1.5]    (488,245.74) .. controls (451.2,224.77) and (436,215.74) .. (413,221.74) ;
\draw [shift={(446,223.88)}, rotate = 22.36] [fill={rgb, 255:red, 0; green, 0; blue, 0 }  ][line width=0.08]  [draw opacity=0] (11.61,-5.58) -- (0,0) -- (11.61,5.58) -- cycle    ;
%Curve Lines [id:da4347917969226498] 
\draw [line width=1.5]    (322,102.74) .. controls (380.51,117.74) and (405.2,99.44) .. (476.53,71.44) ;
\draw [shift={(408.17,98.87)}, rotate = 161.31] [fill={rgb, 255:red, 0; green, 0; blue, 0 }  ][line width=0.08]  [draw opacity=0] (11.61,-5.58) -- (0,0) -- (11.61,5.58) -- cycle    ;
%Curve Lines [id:da8431189883874404] 
\draw [color={rgb, 255:red, 0; green, 0; blue, 0 }  ,draw opacity=1 ][line width=1.5]    (200.49,65.74) .. controls (273,45.74) and (248.51,90.74) .. (305,97.74) ;
\draw [shift={(261.13,73.98)}, rotate = 224.04] [fill={rgb, 255:red, 0; green, 0; blue, 0 }  ,fill opacity=1 ][line width=0.08]  [draw opacity=0] (11.61,-5.58) -- (0,0) -- (11.61,5.58) -- cycle    ;
%Curve Lines [id:da2836962945567163] 
\draw [line width=1.5]    (56.53,76.77) .. controls (104.53,52.51) and (130.49,98.47) .. (188.49,67.74) ;
\draw [shift={(129.58,76.99)}, rotate = 191.43] [fill={rgb, 255:red, 0; green, 0; blue, 0 }  ][line width=0.08]  [draw opacity=0] (11.61,-5.58) -- (0,0) -- (11.61,5.58) -- cycle    ;
%Image [id:dp9421059015303361] 
\draw (193.8,66.01) node  {\includegraphics[width=31.7pt,height=26.98pt]{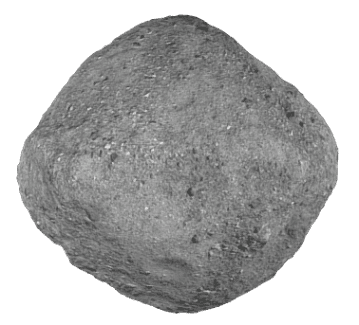}};
%Image [id:dp33406624615743086] 
\draw (490.4,67.65) node  {\includegraphics[width=29.4pt,height=26.48pt]{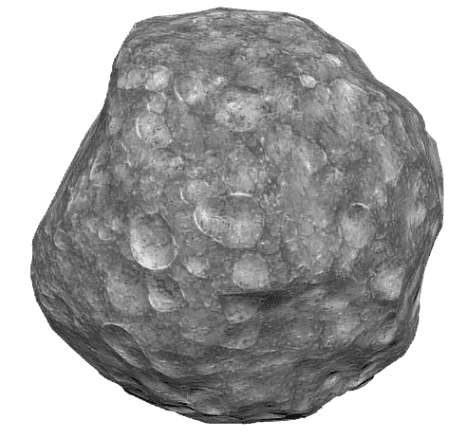}};
%Image [id:dp15492634615252654] 
\draw (404.67,223.65) node  {\includegraphics[width=29.4pt,height=26.48pt]{phoebe.png}};
%Image [id:dp1236188941863905] 
\draw (218.87,242.01) node  {\includegraphics[width=31.7pt,height=26.98pt]{bennu.png}};
%Image [id:dp3992638967517692] 
\draw (314.13,99.32) node  {\includegraphics[width=29.2pt,height=25.02pt]{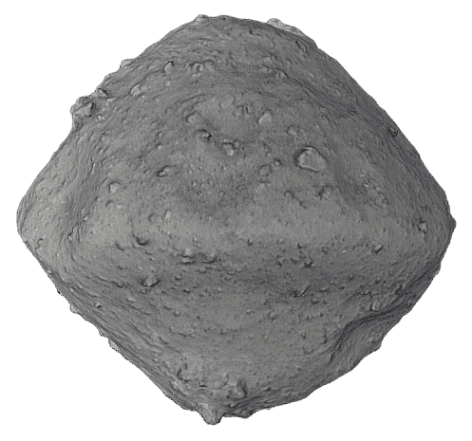}};
%Image [id:dp5709489846071427] 
\draw (491.47,242.65) node  {\includegraphics[width=29.2pt,height=25.02pt]{ryugu.png}};
%Image [id:dp007849710433697643] 
\draw (50,80) node  {\includegraphics[width=45pt,height=45pt]{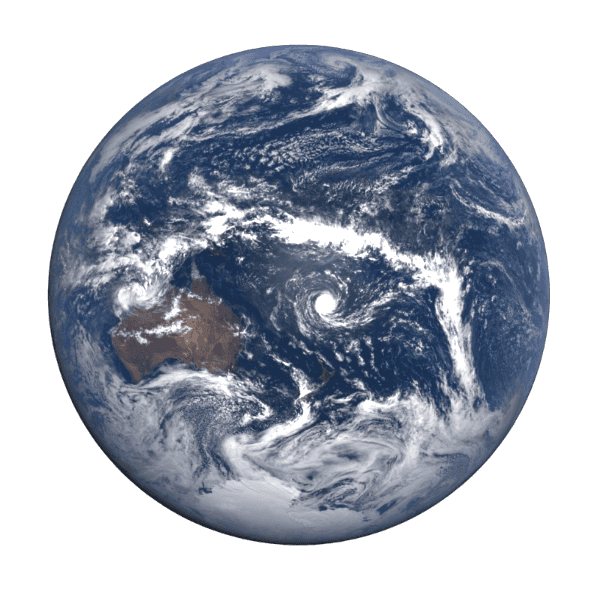}};
%Image [id:dp6493924565066118] 
\draw (50,230) node  {\includegraphics[width=45pt,height=45pt]{earth.png}};

% Text Node
\draw (345.24,21) node  [font=\normalsize] [align=left] {\begin{minipage}[lt]{264.87pt}\setlength\topsep{0pt}
\begin{center}
\textbf{miner deployment phase}
\end{center}

\end{minipage}};
% Text Node
\draw (345.24,169) node  [font=\normalsize] [align=left] {\begin{minipage}[lt]{264.87pt}\setlength\topsep{0pt}
\begin{center}
\textbf{mass collection phase}
\end{center}

\end{minipage}};
% Text Node
\draw (50,126) node   [align=left] {\begin{minipage}[lt]{68pt}\setlength\topsep{0pt}
\begin{center}
Earth departure\\MJD $\displaystyle \geq 64328$
\end{center}

\end{minipage}};
% Text Node
\draw (51.33,276) node   [align=left] {\begin{minipage}[lt]{68pt}\setlength\topsep{0pt}
\begin{center}
Earth return\\MJD $\displaystyle \leq 69807$
\end{center}

\end{minipage}};
% Text Node
\draw (52,189.67) node   [align=left] {\begin{minipage}[lt]{54.4pt}\setlength\topsep{0pt}
\begin{center}
$\displaystyle v_{\infty }$$\displaystyle \leq 6$ km/s
\end{center}

\end{minipage}};
% Text Node
\draw (50.67,41.17) node   [align=left] {\begin{minipage}[lt]{54.4pt}\setlength\topsep{0pt}
\begin{center}
$\displaystyle v_{\infty }$$\displaystyle \leq 6$ km/s
\end{center}

\end{minipage}};
% Text Node
\draw (195.33,92.33) node  [font=\small] [align=left] {\begin{minipage}[lt]{68pt}\setlength\topsep{0pt}
\begin{center}
deployment 1
\end{center}

\end{minipage}};
% Text Node
\draw (313.33,74.33) node  [font=\small] [align=left] {\begin{minipage}[lt]{68pt}\setlength\topsep{0pt}
\begin{center}
deployment 2
\end{center}

\end{minipage}};
% Text Node
\draw (492,93.67) node  [font=\small] [align=left] {\begin{minipage}[lt]{68pt}\setlength\topsep{0pt}
\begin{center}
deployment 3
\end{center}

\end{minipage}};
% Text Node
\draw (218,217) node  [font=\small] [align=left] {\begin{minipage}[lt]{68pt}\setlength\topsep{0pt}
\begin{center}
collection 3
\end{center}

\end{minipage}};
% Text Node
\draw (405.33,249) node  [font=\small] [align=left] {\begin{minipage}[lt]{68pt}\setlength\topsep{0pt}
\begin{center}
collection 2
\end{center}

\end{minipage}};
% Text Node
\draw (492,218.33) node  [font=\small] [align=left] {\begin{minipage}[lt]{68pt}\setlength\topsep{0pt}
\begin{center}
collection 1
\end{center}

\end{minipage}};
% Text Node
\draw (546.94,156) node  [font=\normalsize] [align=left] {\begin{minipage}[lt]{61.29pt}\setlength\topsep{0pt}
\begin{center}
\textbf{intermediate}\\\textbf{ phase}
\end{center}

\end{minipage}};

\end{tikzpicture}

%% file: figures/flowchart.tex
\tikzset{every picture/.style={line width=0.75pt}} %set default line width to 0.75pt        

\begin{tikzpicture}[x=0.75pt,y=0.75pt,yscale=-1,xscale=1]
%uncomment if require: \path (0,559); %set diagram left start at 0, and has height of 559

%Rounded Rect [id:dp24790387129375202] 
\draw  [dash pattern={on 3.75pt off 3pt on 7.5pt off 1.5pt}][line width=1.5]  (0,74) .. controls (0,66.27) and (6.27,60) .. (14,60) -- (196,60) .. controls (203.73,60) and (210,66.27) .. (210,74) -- (210,306) .. controls (210,313.73) and (203.73,320) .. (196,320) -- (14,320) .. controls (6.27,320) and (0,313.73) .. (0,306) -- cycle ;
%Rounded Rect [id:dp03393841514592766] 
\draw  [dash pattern={on 3.75pt off 3pt on 7.5pt off 1.5pt}][line width=1.5]  (410,74) .. controls (410,66.27) and (416.27,60) .. (424,60) -- (606,60) .. controls (613.73,60) and (620,66.27) .. (620,74) -- (620,306) .. controls (620,313.73) and (613.73,320) .. (606,320) -- (424,320) .. controls (416.27,320) and (410,313.73) .. (410,306) -- cycle ;
%Flowchart: Alternative Process [id:dp0904098751172473] 
\draw  [fill={rgb, 255:red, 78; green, 121; blue, 167 }  ,fill opacity=0.5 ][line width=0.75]  (10,87) .. controls (10,83.13) and (13.13,80) .. (17,80) -- (193,80) .. controls (196.87,80) and (200,83.13) .. (200,87) -- (200,113) .. controls (200,116.87) and (196.87,120) .. (193,120) -- (17,120) .. controls (13.13,120) and (10,116.87) .. (10,113) -- cycle ;
%Flowchart: Alternative Process [id:dp30418373033261337] 
\draw  [fill={rgb, 255:red, 78; green, 121; blue, 167 }  ,fill opacity=0.5 ] (10,147) .. controls (10,143.13) and (13.13,140) .. (17,140) -- (193,140) .. controls (196.87,140) and (200,143.13) .. (200,147) -- (200,173) .. controls (200,176.87) and (196.87,180) .. (193,180) -- (17,180) .. controls (13.13,180) and (10,176.87) .. (10,173) -- cycle ;
%Flowchart: Alternative Process [id:dp4900505989421955] 
\draw  [fill={rgb, 255:red, 78; green, 121; blue, 167 }  ,fill opacity=0.5 ] (10,207) .. controls (10,203.13) and (13.13,200) .. (17,200) -- (193,200) .. controls (196.87,200) and (200,203.13) .. (200,207) -- (200,233) .. controls (200,236.87) and (196.87,240) .. (193,240) -- (17,240) .. controls (13.13,240) and (10,236.87) .. (10,233) -- cycle ;
%Flowchart: Alternative Process [id:dp8635320719696251] 
\draw  [fill={rgb, 255:red, 242; green, 142; blue, 43 }  ,fill opacity=0.5 ] (10,267) .. controls (10,263.13) and (13.13,260) .. (17,260) -- (193,260) .. controls (196.87,260) and (200,263.13) .. (200,267) -- (200,293) .. controls (200,296.87) and (196.87,300) .. (193,300) -- (17,300) .. controls (13.13,300) and (10,296.87) .. (10,293) -- cycle ;
%Flowchart: Alternative Process [id:dp5998725586966571] 
\draw  [fill={rgb, 255:red, 78; green, 121; blue, 167 }  ,fill opacity=0.5 ] (420,87) .. controls (420,83.13) and (423.13,80) .. (427,80) -- (603,80) .. controls (606.87,80) and (610,83.13) .. (610,87) -- (610,113) .. controls (610,116.87) and (606.87,120) .. (603,120) -- (427,120) .. controls (423.13,120) and (420,116.87) .. (420,113) -- cycle ;
%Flowchart: Alternative Process [id:dp22809726183408507] 
\draw  [fill={rgb, 255:red, 78; green, 121; blue, 167 }  ,fill opacity=0.5 ] (420,147) .. controls (420,143.13) and (423.13,140) .. (427,140) -- (603,140) .. controls (606.87,140) and (610,143.13) .. (610,147) -- (610,173) .. controls (610,176.87) and (606.87,180) .. (603,180) -- (427,180) .. controls (423.13,180) and (420,176.87) .. (420,173) -- cycle ;
%Flowchart: Alternative Process [id:dp9712344530074639] 
\draw  [fill={rgb, 255:red, 78; green, 121; blue, 167 }  ,fill opacity=0.5 ] (420,207) .. controls (420,203.13) and (423.13,200) .. (427,200) -- (603,200) .. controls (606.87,200) and (610,203.13) .. (610,207) -- (610,233) .. controls (610,236.87) and (606.87,240) .. (603,240) -- (427,240) .. controls (423.13,240) and (420,236.87) .. (420,233) -- cycle ;
%Flowchart: Alternative Process [id:dp10398502503972873] 
\draw  [fill={rgb, 255:red, 242; green, 142; blue, 43 }  ,fill opacity=0.5 ] (420,267) .. controls (420,263.13) and (423.13,260) .. (427,260) -- (603,260) .. controls (606.87,260) and (610,263.13) .. (610,267) -- (610,293) .. controls (610,296.87) and (606.87,300) .. (603,300) -- (427,300) .. controls (423.13,300) and (420,296.87) .. (420,293) -- cycle ;
%Flowchart: Alternative Process [id:dp385336092941456] 
\draw  [fill={rgb, 255:red, 118; green, 183; blue, 178 }  ,fill opacity=0.5 ] (250,75.25) .. controls (250,72.35) and (252.35,70) .. (255.25,70) -- (364.75,70) .. controls (367.65,70) and (370,72.35) .. (370,75.25) -- (370,94.75) .. controls (370,97.65) and (367.65,100) .. (364.75,100) -- (255.25,100) .. controls (252.35,100) and (250,97.65) .. (250,94.75) -- cycle ;
%Flowchart: Alternative Process [id:dp8555710925373048] 
\draw  [fill={rgb, 255:red, 242; green, 142; blue, 43 }  ,fill opacity=0.5 ] (260,295.25) .. controls (260,292.35) and (262.35,290) .. (265.25,290) -- (354.75,290) .. controls (357.65,290) and (360,292.35) .. (360,295.25) -- (360,314.75) .. controls (360,317.65) and (357.65,320) .. (354.75,320) -- (265.25,320) .. controls (262.35,320) and (260,317.65) .. (260,314.75) -- cycle ;
%Flowchart: Alternative Process [id:dp2741611515460336] 
\draw  [fill={rgb, 255:red, 118; green, 183; blue, 178 }  ,fill opacity=0.5 ] (250,117) .. controls (250,113.13) and (253.13,110) .. (257,110) -- (363,110) .. controls (366.87,110) and (370,113.13) .. (370,117) -- (370,143) .. controls (370,146.87) and (366.87,150) .. (363,150) -- (257,150) .. controls (253.13,150) and (250,146.87) .. (250,143) -- cycle ;
%Rounded Rect [id:dp9954424134667028] 
\draw  [dash pattern={on 3.75pt off 3pt on 7.5pt off 1.5pt}][line width=1.5]  (240,66.67) .. controls (240,62.98) and (242.98,60) .. (246.67,60) -- (373.33,60) .. controls (377.02,60) and (380,62.98) .. (380,66.67) -- (380,153.33) .. controls (380,157.02) and (377.02,160) .. (373.33,160) -- (246.67,160) .. controls (242.98,160) and (240,157.02) .. (240,153.33) -- cycle ;
%Straight Lines [id:da14878023191757817] 
\draw [line width=1.5]    (210,190) -- (406,190) ;
\draw [shift={(410,190)}, rotate = 180] [fill={rgb, 255:red, 0; green, 0; blue, 0 }  ][line width=0.08]  [draw opacity=0] (9.29,-4.46) -- (0,0) -- (9.29,4.46) -- cycle    ;
%Straight Lines [id:da6386418037883479] 
\draw [line width=1.5]    (240,110) -- (214,110) ;
\draw [shift={(210,110)}, rotate = 360] [fill={rgb, 255:red, 0; green, 0; blue, 0 }  ][line width=0.08]  [draw opacity=0] (9.29,-4.46) -- (0,0) -- (9.29,4.46) -- cycle    ;
%Straight Lines [id:da23026424856739425] 
\draw [line width=1.5]    (410,240) -- (374,240) ;
\draw [shift={(370,240)}, rotate = 360] [fill={rgb, 255:red, 0; green, 0; blue, 0 }  ][line width=0.08]  [draw opacity=0] (9.29,-4.46) -- (0,0) -- (9.29,4.46) -- cycle    ;
%Straight Lines [id:da22513051842317755] 
\draw [line width=1.5]    (250,240) -- (214,240) ;
\draw [shift={(210,240)}, rotate = 360] [fill={rgb, 255:red, 0; green, 0; blue, 0 }  ][line width=0.08]  [draw opacity=0] (9.29,-4.46) -- (0,0) -- (9.29,4.46) -- cycle    ;
%Flowchart: Preparation [id:dp044802623149849286] 
\draw  [fill={rgb, 255:red, 225; green, 87; blue, 89 }  ,fill opacity=0.5 ][line width=1.5]  (250,240) -- (272.5,220) -- (347.5,220) -- (370,240) -- (347.5,260) -- (272.5,260) -- cycle ;
%Straight Lines [id:da22430652069518286] 
\draw [line width=1.5]    (310,260) -- (310,286) ;
\draw [shift={(310,290)}, rotate = 270] [fill={rgb, 255:red, 0; green, 0; blue, 0 }  ][line width=0.08]  [draw opacity=0] (9.29,-4.46) -- (0,0) -- (9.29,4.46) -- cycle    ;
%Straight Lines [id:da20688839581373641] 
\draw [line width=1.5]    (105,60) -- (105,76) ;
\draw [shift={(105,80)}, rotate = 270] [fill={rgb, 255:red, 0; green, 0; blue, 0 }  ][line width=0.08]  [draw opacity=0] (9.29,-4.46) -- (0,0) -- (9.29,4.46) -- cycle    ;
%Straight Lines [id:da15787869416365385] 
\draw [line width=1.5]    (105,120) -- (105,136) ;
\draw [shift={(105,140)}, rotate = 270] [fill={rgb, 255:red, 0; green, 0; blue, 0 }  ][line width=0.08]  [draw opacity=0] (9.29,-4.46) -- (0,0) -- (9.29,4.46) -- cycle    ;
%Straight Lines [id:da49059660442547326] 
\draw [line width=1.5]    (105,180) -- (105,196) ;
\draw [shift={(105,200)}, rotate = 270] [fill={rgb, 255:red, 0; green, 0; blue, 0 }  ][line width=0.08]  [draw opacity=0] (9.29,-4.46) -- (0,0) -- (9.29,4.46) -- cycle    ;
%Straight Lines [id:da08137709175142405] 
\draw [line width=1.5]    (105,240) -- (105,256) ;
\draw [shift={(105,260)}, rotate = 270] [fill={rgb, 255:red, 0; green, 0; blue, 0 }  ][line width=0.08]  [draw opacity=0] (9.29,-4.46) -- (0,0) -- (9.29,4.46) -- cycle    ;
%Straight Lines [id:da18979402204847662] 
\draw [line width=1.5]    (105,300) -- (105,316) ;
\draw [shift={(105,320)}, rotate = 270] [fill={rgb, 255:red, 0; green, 0; blue, 0 }  ][line width=0.08]  [draw opacity=0] (9.29,-4.46) -- (0,0) -- (9.29,4.46) -- cycle    ;
%Straight Lines [id:da658232190014171] 
\draw [line width=1.5]    (515,60) -- (515,76) ;
\draw [shift={(515,80)}, rotate = 270] [fill={rgb, 255:red, 0; green, 0; blue, 0 }  ][line width=0.08]  [draw opacity=0] (9.29,-4.46) -- (0,0) -- (9.29,4.46) -- cycle    ;
%Straight Lines [id:da903010017364497] 
\draw [line width=1.5]    (515,120) -- (515,136) ;
\draw [shift={(515,140)}, rotate = 270] [fill={rgb, 255:red, 0; green, 0; blue, 0 }  ][line width=0.08]  [draw opacity=0] (9.29,-4.46) -- (0,0) -- (9.29,4.46) -- cycle    ;
%Straight Lines [id:da10824049938917901] 
\draw [line width=1.5]    (500,180) -- (500,196) ;
\draw [shift={(500,200)}, rotate = 270] [fill={rgb, 255:red, 0; green, 0; blue, 0 }  ][line width=0.08]  [draw opacity=0] (9.29,-4.46) -- (0,0) -- (9.29,4.46) -- cycle    ;
%Straight Lines [id:da4417134005822232] 
\draw [line width=1.5]    (515,240) -- (515,256) ;
\draw [shift={(515,260)}, rotate = 270] [fill={rgb, 255:red, 0; green, 0; blue, 0 }  ][line width=0.08]  [draw opacity=0] (9.29,-4.46) -- (0,0) -- (9.29,4.46) -- cycle    ;
%Straight Lines [id:da48311416786691597] 
\draw [line width=1.5]    (515,300) -- (515,316) ;
\draw [shift={(515,320)}, rotate = 270] [fill={rgb, 255:red, 0; green, 0; blue, 0 }  ][line width=0.08]  [draw opacity=0] (9.29,-4.46) -- (0,0) -- (9.29,4.46) -- cycle    ;
%Straight Lines [id:da9421246976799393] 
\draw [line width=1.5]    (530,200) -- (530,184) ;
\draw [shift={(530,180)}, rotate = 90] [fill={rgb, 255:red, 0; green, 0; blue, 0 }  ][line width=0.08]  [draw opacity=0] (9.29,-4.46) -- (0,0) -- (9.29,4.46) -- cycle    ;

% Text Node
\draw (105,45) node  [font=\small] [align=left] {\begin{minipage}[lt]{142.8pt}\setlength\topsep{0pt}
\begin{center}
\textbf{Binary Integer Program (BIP)}
\end{center}

\end{minipage}};
% Text Node
\draw (105,100) node   [align=left] {\begin{minipage}[lt]{129.2pt}\setlength\topsep{0pt}
\begin{center}
Asteroid-to-asteroid transfer cost approximation
\end{center}

\end{minipage}};
% Text Node
\draw (105,160) node   [align=left] {\begin{minipage}[lt]{129.2pt}\setlength\topsep{0pt}
\begin{center}
Pruning of infeasible or costly transfers
\end{center}

\end{minipage}};
% Text Node
\draw (105,220) node   [align=left] {\begin{minipage}[lt]{129.2pt}\setlength\topsep{0pt}
\begin{center}
Formulation and solving of BIP problem
\end{center}

\end{minipage}};
% Text Node
\draw (105,280) node   [align=left] {\begin{minipage}[lt]{129.2pt}\setlength\topsep{0pt}
\begin{center}
Best rendezvous sequences and corresponding times
\end{center}

\end{minipage}};
% Text Node
\draw (515,160) node   [align=left] {\begin{minipage}[lt]{129.2pt}\setlength\topsep{0pt}
\begin{center}
Dynamical linearization around reference trajectory
\end{center}

\end{minipage}};
% Text Node
\draw (515,220) node   [align=left] {\begin{minipage}[lt]{129.2pt}\setlength\topsep{0pt}
\begin{center}
Formulation and solving of convex problem
\end{center}

\end{minipage}};
% Text Node
\draw (515,280) node   [align=left] {\begin{minipage}[lt]{129.2pt}\setlength\topsep{0pt}
\begin{center}
Optimal control profile and optimal rendezvous times
\end{center}

\end{minipage}};
% Text Node
\draw (310,85) node  [font=\normalsize] [align=left] {\begin{minipage}[lt]{81.6pt}\setlength\topsep{0pt}
\begin{center}
Asteroid subset
\end{center}

\end{minipage}};
% Text Node
\draw (310,305) node  [font=\normalsize] [align=left] {\begin{minipage}[lt]{47.6pt}\setlength\topsep{0pt}
\begin{center}
Terminate
\end{center}

\end{minipage}};
% Text Node
\draw (310,130) node  [font=\normalsize] [align=left] {\begin{minipage}[lt]{81.6pt}\setlength\topsep{0pt}
\begin{center}
Rendezvous times approximate guess
\end{center}

\end{minipage}};
% Text Node
\draw (515,45) node  [font=\small] [align=left] {\begin{minipage}[lt]{142.8pt}\setlength\topsep{0pt}
\begin{center}
\textbf{Sequential Convex Program (SCP)}
\end{center}

\end{minipage}};
% Text Node
\draw (310,45) node  [font=\small] [align=left] {\begin{minipage}[lt]{95.2pt}\setlength\topsep{0pt}
\begin{center}
\textbf{Nested-loop Inputs}
\end{center}

\end{minipage}};
% Text Node
\draw (310,240) node   [align=left] {\begin{minipage}[lt]{81.6pt}\setlength\topsep{0pt}
\begin{center}
Improvement \\in objective?
\end{center}

\end{minipage}};
% Text Node
\draw (515,100) node   [align=left] {\begin{minipage}[lt]{129.2pt}\setlength\topsep{0pt}
\begin{center}
Reference trajectory generation
\end{center}

\end{minipage}};
% Text Node
\draw (297,273) node  [font=\normalsize] [align=left] {\begin{minipage}[lt]{27.2pt}\setlength\topsep{0pt}
\begin{center}
No
\end{center}

\end{minipage}};
% Text Node
\draw (232,230.63) node  [font=\normalsize] [align=left] {\begin{minipage}[lt]{27.2pt}\setlength\topsep{0pt}
\begin{center}
Yes
\end{center}

\end{minipage}};

\end{tikzpicture}

%% file: figures/scp_explainer.tex
\tikzset{every picture/.style={line width=0.75pt}} %set default line width to 0.75pt        

\begin{tikzpicture}[x=0.75pt,y=0.75pt,yscale=-1,xscale=1]
%uncomment if require: \path (0,317); %set diagram left start at 0, and has height of 317

%Straight Lines [id:da8681342473264935] 
\draw    (562.08,94.72) -- (567.28,109.12) ;
%Straight Lines [id:da6917483945242333] 
\draw [color={rgb, 255:red, 242; green, 142; blue, 43 }  ,draw opacity=1 ][line width=1.5]    (336.67,101.33) -- (347.77,117.05) ;
\draw [shift={(350.08,120.32)}, rotate = 234.76] [fill={rgb, 255:red, 242; green, 142; blue, 43 }  ,fill opacity=1 ][line width=0.08]  [draw opacity=0] (6.97,-3.35) -- (0,0) -- (6.97,3.35) -- cycle    ;
%Straight Lines [id:da9018492461028766] 
\draw [color={rgb, 255:red, 242; green, 142; blue, 43 }  ,draw opacity=1 ][line width=1.5]    (484.67,96) -- (478.14,104.37) ;
\draw [shift={(475.68,107.52)}, rotate = 307.96] [fill={rgb, 255:red, 242; green, 142; blue, 43 }  ,fill opacity=1 ][line width=0.08]  [draw opacity=0] (6.97,-3.35) -- (0,0) -- (6.97,3.35) -- cycle    ;
%Straight Lines [id:da9094090016320373] 
\draw [color={rgb, 255:red, 242; green, 142; blue, 43 }  ,draw opacity=1 ][line width=1.5]    (426,86) -- (419.95,107.67) ;
\draw [shift={(418.88,111.52)}, rotate = 285.59] [fill={rgb, 255:red, 242; green, 142; blue, 43 }  ,fill opacity=1 ][line width=0.08]  [draw opacity=0] (6.97,-3.35) -- (0,0) -- (6.97,3.35) -- cycle    ;
%Straight Lines [id:da4843718683110665] 
\draw [color={rgb, 255:red, 242; green, 142; blue, 43 }  ,draw opacity=1 ][line width=1.5]    (269.2,112) -- (279.01,95.37) ;
\draw [shift={(281.04,91.92)}, rotate = 120.53] [fill={rgb, 255:red, 242; green, 142; blue, 43 }  ,fill opacity=1 ][line width=0.08]  [draw opacity=0] (6.97,-3.35) -- (0,0) -- (6.97,3.35) -- cycle    ;
%Straight Lines [id:da5145465004814822] 
\draw [color={rgb, 255:red, 242; green, 142; blue, 43 }  ,draw opacity=1 ][line width=1.5]    (216,98) -- (226.11,88.64) ;
\draw [shift={(229.04,85.92)}, rotate = 137.19] [fill={rgb, 255:red, 242; green, 142; blue, 43 }  ,fill opacity=1 ][line width=0.08]  [draw opacity=0] (6.97,-3.35) -- (0,0) -- (6.97,3.35) -- cycle    ;
%Straight Lines [id:da6205649948549914] 
\draw [color={rgb, 255:red, 242; green, 142; blue, 43 }  ,draw opacity=1 ][line width=1.5]    (170.67,82.67) -- (177.53,81.17) ;
\draw [shift={(181.44,80.32)}, rotate = 167.71] [fill={rgb, 255:red, 242; green, 142; blue, 43 }  ,fill opacity=1 ][line width=0.08]  [draw opacity=0] (6.97,-3.35) -- (0,0) -- (6.97,3.35) -- cycle    ;
%Straight Lines [id:da6506001005149267] 
\draw [color={rgb, 255:red, 242; green, 142; blue, 43 }  ,draw opacity=1 ][line width=1.5]    (126,74) -- (145.72,66.19) ;
\draw [shift={(149.44,64.72)}, rotate = 158.4] [fill={rgb, 255:red, 242; green, 142; blue, 43 }  ,fill opacity=1 ][line width=0.08]  [draw opacity=0] (6.97,-3.35) -- (0,0) -- (6.97,3.35) -- cycle    ;
%Straight Lines [id:da15718774475179154] 
\draw [color={rgb, 255:red, 242; green, 142; blue, 43 }  ,draw opacity=1 ][line width=1.5]    (74.67,72.67) -- (90.89,58.11) ;
\draw [shift={(93.87,55.44)}, rotate = 138.1] [fill={rgb, 255:red, 242; green, 142; blue, 43 }  ,fill opacity=1 ][line width=0.08]  [draw opacity=0] (6.97,-3.35) -- (0,0) -- (6.97,3.35) -- cycle    ;
%Straight Lines [id:da014623704836826734] 
\draw [color={rgb, 255:red, 225; green, 87; blue, 89 }  ,draw opacity=1 ][line width=1.5]    (58,81.06) -- (118.94,51.2) ;
\draw [shift={(122.53,49.44)}, rotate = 153.9] [fill={rgb, 255:red, 225; green, 87; blue, 89 }  ,fill opacity=1 ][line width=0.08]  [draw opacity=0] (11.61,-5.58) -- (0,0) -- (11.61,5.58) -- cycle    ;
%Curve Lines [id:da8431189883874404] 
\draw [color={rgb, 255:red, 0; green, 0; blue, 0 }  ,draw opacity=1 ][line width=1.5]    (323.4,105.6) .. controls (451.8,56.2) and (484.4,118.6) .. (570,100) ;
%Curve Lines [id:da2836962945567163] 
\draw [line width=1.5]    (56.53,76.77) .. controls (157.4,50.6) and (259.4,137.34) .. (317.4,106.6) ;
%Image [id:dp9421059015303361] 
\draw (318.87,107.99) node  {\includegraphics[width=31.7pt,height=26.98pt]{bennu.png}};
%Image [id:dp3992638967517692] 
\draw (588.1,99.12) node  {\includegraphics[width=37.95pt,height=32.52pt]{ryugu.png}};
%Image [id:dp007849710433697643] 
\draw (50,80) node  {\includegraphics[width=45pt,height=45pt]{earth.png}};
%Straight Lines [id:da3914597456382478] 
\draw    (73.33,66.33) -- (75.87,78.77) ;
%Straight Lines [id:da5471718110097761] 
\draw    (126.27,67.96) -- (127.04,81.36) ;
%Straight Lines [id:da469964562926136] 
\draw    (171.04,76.96) -- (167.84,90.96) ;
%Straight Lines [id:da172519646922501] 
\draw    (217.84,88.96) -- (215.04,104.96) ;
%Straight Lines [id:da05308601597906404] 
\draw    (270.24,104.16) -- (269.04,118.16) ;
%Straight Lines [id:da09170934823087284] 
\draw    (333.44,93.36) -- (339.44,109.76) ;
%Straight Lines [id:da6143132910617306] 
\draw    (426.88,78.56) -- (425.44,94.56) ;
%Straight Lines [id:da6697395682578182] 
\draw    (485.28,89.92) -- (484.48,103.36) ;
%Straight Lines [id:da30647620929744823] 
\draw    (299.04,104.96) -- (302.08,118.72) ;

% Text Node
\draw (101.12,89.7) node  [font=\small] [align=left] {\begin{minipage}[lt]{31.66pt}\setlength\topsep{0pt}
\begin{center}
$\displaystyle \Delta t_{1,1}$
\end{center}

\end{minipage}};
% Text Node
\draw (147.39,98.17) node  [font=\small] [align=left] {\begin{minipage}[lt]{31.66pt}\setlength\topsep{0pt}
\begin{center}
$\displaystyle \Delta t_{1,2}$
\end{center}

\end{minipage}};
% Text Node
\draw (187.92,113.3) node  [font=\small] [align=left] {\begin{minipage}[lt]{31.66pt}\setlength\topsep{0pt}
\begin{center}
$\displaystyle \Delta t_{1,3}$
\end{center}

\end{minipage}};
% Text Node
\draw (238.6,126.9) node  [font=\small] [align=left] {\begin{minipage}[lt]{31.66pt}\setlength\topsep{0pt}
\begin{center}
$\displaystyle \Delta t_{1,4}$
\end{center}

\end{minipage}};
% Text Node
\draw (286.6,136.9) node  [font=\small] [align=left] {\begin{minipage}[lt]{31.66pt}\setlength\topsep{0pt}
\begin{center}
$\displaystyle \Delta t_{1,5}$
\end{center}

\end{minipage}};
% Text Node
\draw (377.92,72.5) node  [font=\small] [align=left] {\begin{minipage}[lt]{31.66pt}\setlength\topsep{0pt}
\begin{center}
$\displaystyle \Delta t_{2,1}$
\end{center}

\end{minipage}};
% Text Node
\draw (461.12,77.3) node  [font=\small] [align=left] {\begin{minipage}[lt]{31.66pt}\setlength\topsep{0pt}
\begin{center}
$\displaystyle \Delta t_{2,2}$
\end{center}

\end{minipage}};
% Text Node
\draw (527,86.5) node  [font=\small] [align=left] {\begin{minipage}[lt]{31.66pt}\setlength\topsep{0pt}
\begin{center}
$\displaystyle \Delta t_{2,3}$
\end{center}

\end{minipage}};
% Text Node
\draw (51,119.6) node  [font=\normalsize] [align=left] {\begin{minipage}[lt]{31.66pt}\setlength\topsep{0pt}
\begin{center}
$\displaystyle t_{1}$
\end{center}

\end{minipage}};
% Text Node
\draw (50.58,32.6) node  [font=\small] [align=left] {\begin{minipage}[lt]{41.21pt}\setlength\topsep{0pt}
\begin{center}
Earth\\departure
\end{center}

\end{minipage}};
% Text Node
\draw (317.7,72.43) node  [font=\small] [align=left] {\begin{minipage}[lt]{43.08pt}\setlength\topsep{0pt}
\begin{center}
asteroid 1 rendezvous
\end{center}

\end{minipage}};
% Text Node
\draw (587.3,60.83) node  [font=\small] [align=left] {\begin{minipage}[lt]{43.08pt}\setlength\topsep{0pt}
\begin{center}
asteroid 2 rendezvous
\end{center}

\end{minipage}};
% Text Node
\draw (320.6,139.2) node  [font=\normalsize] [align=left] {\begin{minipage}[lt]{31.66pt}\setlength\topsep{0pt}
\begin{center}
$\displaystyle t_{2}$
\end{center}

\end{minipage}};
% Text Node
\draw (589.4,133.2) node  [font=\normalsize] [align=left] {\begin{minipage}[lt]{31.66pt}\setlength\topsep{0pt}
\begin{center}
$\displaystyle t_{3}$
\end{center}

\end{minipage}};
% Text Node
\draw (140.72,43.46) node  [font=\small] [align=left] {\begin{minipage}[lt]{31.66pt}\setlength\topsep{0pt}
\begin{center}
$\displaystyle v_{1,\text{dep}}$
\end{center}

\end{minipage}};

\end{tikzpicture}

%% file: main.bbl
\begin{thebibliography}{10}
\expandafter\ifx\csname url\endcsname\relax
  \def\url#1{\texttt{#1}}\fi
\expandafter\ifx\csname urlprefix\endcsname\relax\def\urlprefix{URL }\fi
\expandafter\ifx\csname href\endcsname\relax
  \def\href#1#2{#2} \def\path#1{#1}\fi

\bibitem{hallPioneer10111974}
C.~F. Hall, Pioneer 10/11 spacecraft and missions to {{Jupiter}}, in: International {{Astronautical Federation}}, {{International Astronautical Congress}}, Amsterdam, 1974, p.~24.

\bibitem{reinhardGiottoEncounterComet1986}
R.~Reinhard, The {{Giotto}} encounter with comet {{Halley}}, Nature 321~(6067) (1986) 313--318.
\newblock \href {https://doi.org/10.1038/321313a0} {\path{doi:10.1038/321313a0}}.

\bibitem{schmidtMarsExpressESA2003a}
R.~Schmidt, Mars {{Express}}---{{ESA}}'s first mission to planet {{Mars}}, Acta Astronautica 52~(2) (2003) 197--202.
\newblock \href {https://doi.org/10.1016/S0094-5765(02)00157-1} {\path{doi:10.1016/S0094-5765(02)00157-1}}.

\bibitem{grassetJUpiterICyMoons2013}
O.~Grasset, M.~K. Dougherty, A.~Coustenis, E.~J. Bunce, C.~Erd, D.~Titov, M.~Blanc, A.~Coates, P.~Drossart, L.~N. Fletcher, H.~Hussmann, R.~Jaumann, N.~Krupp, J.~P. Lebreton, O.~{Prieto-Ballesteros}, P.~Tortora, F.~Tosi, T.~Van~Hoolst, {{JUpiter ICy}} moons {{Explorer}} ({{JUICE}}): {{An ESA}} mission to orbit {{Ganymede}} and to characterise the {{Jupiter}} system, Planetary and Space Science 78 (2013) 1--21.
\newblock \href {https://doi.org/10.1016/j.pss.2012.12.002} {\path{doi:10.1016/j.pss.2012.12.002}}.

\bibitem{englanderTrajectoryDesignLucy2019}
J.~A. Englander, K.~Berry, B.~Sutter, D.~Stanbridge, D.~H. Ellison, K.~Williams, J.~McAdams, J.~M. Knittel, C.~Welch, H.~Levison, Trajectory {{Design}} of the {{Lucy Mission}} to {{Explore}} the {{Diversity}} of the {{Jupiter Trojans}}, 70th International Astronautical Congress (2019).

\bibitem{izzoAutomatedAsteroidSelection2007}
D.~Izzo, T.~Vinko, C.~Bombardelli, S.~Brendelberger, S.~Centuori, Automated asteroid selection for a 'grand tour' mission, 58th International Astronautical Congress (Sep. 2007).

\bibitem{longStochasticMissionExploration2024}
C.~J. Long, D.~Lujan, A.~{Pedros-Faura}, J.~W. McMahon, Stochastic {{Mission Exploration Tool}} for {{Asteroid Tours}}, Journal of Spacecraft and Rockets 0~(0) (2024) 1--14.
\newblock \href {https://doi.org/10.2514/1.A36051} {\path{doi:10.2514/1.A36051}}.

\bibitem{dicarloLowthrustTourMain2018}
M.~Di~Carlo, M.~Vasile, J.~Dunlop, Low-thrust tour of the main belt asteroids, Advances in Space Research 62~(8) (2018) 2026--2045.
\newblock \href {https://doi.org/10.1016/j.asr.2017.12.033} {\path{doi:10.1016/j.asr.2017.12.033}}.

\bibitem{izzo1stACTGlobal2007}
D.~Izzo, 1st {{ACT}} global trajectory optimisation competition: {{Problem}} description and summary of the results, Acta Astronautica 61~(9) (2007) 731--734.
\newblock \href {https://doi.org/10.1016/j.actaastro.2007.03.003} {\path{doi:10.1016/j.actaastro.2007.03.003}}.

\bibitem{schlueterNonlinearMixedInteger2012}
M.~Schlueter, Nonlinear mixed integer based {{Optimization Technique}} for {{Space Applications}}, Ph.D. thesis, The University of Birmingham (May 2012).

\bibitem{rossHybridOptimalControl2005}
I.~M. Ross, C.~N. D'Souza, Hybrid {{Optimal Control Framework}} for {{Mission Planning}}, Journal of Guidance, Control, and Dynamics 28~(4) (2005) 686--697.
\newblock \href {https://doi.org/10.2514/1.8285} {\path{doi:10.2514/1.8285}}.

\bibitem{dambrosioMixedIntegerNonlinear2013}
C.~D'Ambrosio, A.~Lodi, Mixed integer nonlinear programming tools: An updated practical overview, Annals of Operations Research 204~(1) (2013) 301--320.
\newblock \href {https://doi.org/10.1007/s10479-012-1272-5} {\path{doi:10.1007/s10479-012-1272-5}}.

\bibitem{englanderAutomatedMissionPlanning2012}
J.~A. Englander, B.~A. Conway, T.~Williams, Automated {{Mission Planning}} via {{Evolutionary Algorithms}}, Journal of Guidance, Control, and Dynamics 35~(6) (2012) 1878--1887.
\newblock \href {https://doi.org/10.2514/1.54101} {\path{doi:10.2514/1.54101}}.

\bibitem{lawlerTravelingSalesmanProblem1985}
E.~L. Lawler, J.~K. Lenstra, A.~H. G.~R. Kan, D.~B. Shmoys, The {{Traveling Salesman Problem}}, John Wiley \& Sons, Incorporated, 1985.

\bibitem{cormenIntroductionAlgorithmsFourth2022}
T.~H. Cormen, C.~E. Leiserson, R.~L. Rivest, C.~Stein, Introduction to {{Algorithms}}, Fourth Edition, MIT Press, 2022.

\bibitem{lawlerBranchandBoundMethodsSurvey1966}
E.~L. Lawler, D.~E. Wood, Branch-and-{{Bound Methods}}: {{A Survey}}, Operations Research 14~(4) (1966) 699--719.
\newblock \href {https://doi.org/10.1287/opre.14.4.699} {\path{doi:10.1287/opre.14.4.699}}.

\bibitem{padbergBranchCutAlgorithmResolution1991}
M.~Padberg, G.~Rinaldi, A {{Branch-and-Cut Algorithm}} for the {{Resolution}} of {{Large-Scale Symmetric Traveling Salesman Problems}}, SIAM Review 33~(1) (1991) 60--100.
\newblock \href {https://doi.org/10.1137/1033004} {\path{doi:10.1137/1033004}}.

\bibitem{anandComparativeAnalysisOptimization2017}
R.~Anand, D.~Aggarwal, V.~Chahar, A {{Comparative Analysis}} of {{Optimization Solvers}}, Journal of Statistics and Management Systems (Jul. 2017).
\newblock \href {https://doi.org/10.1080/09720510.2017.1395182} {\path{doi:10.1080/09720510.2017.1395182}}.

\bibitem{zhangGTOC11Results2023}
Z.~Zhang, N.~Zhang, X.~Guo, D.~Wu, X.~Xie, J.~Li, J.~Yang, S.~Chen, F.~Jiang, H.~Baoyin, H.~Li, H.~Zheng, X.~Duan, {{GTOC}} 11: {{Results}} from {{Tsinghua University}} and {{Shanghai Institute}} of {{Satellite Engineering}}, Acta Astronautica 202 (2023) 819--828.
\newblock \href {https://doi.org/10.1016/j.actaastro.2022.06.028} {\path{doi:10.1016/j.actaastro.2022.06.028}}.

\bibitem{bellomeTrajectoryDesignMultiTarget2022}
A.~Bellome, Trajectory {{Design}} of {{Multi-Target Missions}} via {{Graph Transcription}} and {{Dynamic Programming}}, Ph.D. thesis, Cranfield University (2022).

\bibitem{blumMetaheuristicsCombinatorialOptimization2003}
C.~Blum, A.~Roli, Metaheuristics in combinatorial optimization: {{Overview}} and conceptual comparison, ACM Comput. Surv. 35~(3) (2003) 268--308.
\newblock \href {https://doi.org/10.1145/937503.937505} {\path{doi:10.1145/937503.937505}}.

\bibitem{conwaySpacecraftTrajectoryOptimization2010}
B.~Conway, Spacecraft {{Trajectory Optimization}}, Cambridge University Press, 2010.
\newblock \href {https://doi.org/10.1017/CBO9780511778025} {\path{doi:10.1017/CBO9780511778025}}.

\bibitem{chaiReviewOptimizationTechniques2019}
R.~Chai, A.~Savvaris, A.~Tsourdos, S.~Chai, Y.~Xia, A review of optimization techniques in spacecraft flight trajectory design, Progress in Aerospace Sciences 109 (2019) 100543.
\newblock \href {https://doi.org/10.1016/j.paerosci.2019.05.003} {\path{doi:10.1016/j.paerosci.2019.05.003}}.

\bibitem{benedikterConvexApproachThreeDimensional2021}
B.~Benedikter, A.~Zavoli, G.~Colasurdo, S.~Pizzurro, E.~Cavallini, Convex {{Approach}} to {{Three-Dimensional Launch Vehicle Ascent Trajectory Optimization}}, Journal of Guidance, Control, and Dynamics 44~(6) (2021) 1116--1131.
\newblock \href {https://doi.org/10.2514/1.G005376} {\path{doi:10.2514/1.G005376}}.

\bibitem{kwonSequentialConvexProgramming2021}
D.~Kwon, Y.~Jung, Y.-J. Cheon, H.~Bang, Sequential convex programming approach for real-time guidance during the powered descent phase of mars landing missions, Advances in Space Research 68~(11) (2021) 4398--4417.
\newblock \href {https://doi.org/10.1016/j.asr.2021.08.033} {\path{doi:10.1016/j.asr.2021.08.033}}.

\bibitem{hofmannComputationalGuidanceLowThrust2023a}
C.~Hofmann, Computational {{Guidance}} for {{Low-Thrust Spacecraft}} in {{Deep Space Based}} on {{Convex Optimization}}, Ph.D. thesis, Politecnico di Milano (2023).

\bibitem{kumagaiAdaptiveMeshSequentialConvex2024}
N.~Kumagai, K.~Oguri, Adaptive-{{Mesh Sequential Convex Programming}} for {{Space Trajectory Optimization}}, Journal of Guidance, Control, and Dynamics 0~(0) (2024) 1--8.
\newblock \href {https://doi.org/10.2514/1.G008107} {\path{doi:10.2514/1.G008107}}.

\bibitem{ahujaNetworkFlowsTheory1993}
R.~K. Ahuja, T.~L. Magnanti, J.~B. Orlin, Network Flows: Theory, Algorithms, and Applications, Prentice Hall, Upper Saddle River, NJ, 1993.

\bibitem{gtoc12ProblemDescription}
{Z. Zhang, N. Zhang, X. Guo, D. Wu, X. Xie, J. Yang, F. Jiang, H. Baoyin}, {Sustainable Asteroid Mining: on the design of GTOC12 problem and summary of the results}, Astrodynamics (2024).

\bibitem{armellinGTOC12Results2024}
R.~Armellin, A.~Bellome, X.~Fu, H.~Holt, C.~Parigini, M.~Wijayatunga, J.~Yarndley, {{GTOC}} 12: {{Results}} from {{TheAntipodes}}, Astrodynamics (accepted for publication, 2024).

\bibitem{lubinJuMPRecentImprovements2023}
M.~Lubin, O.~Dowson, J.~D. Garcia, J.~Huchette, B.~Legat, J.~P. Vielma, {{JuMP}} 1.0: Recent improvements to a modeling language for mathematical optimization, Mathematical Programming Computation 15~(3) (2023) 581--589.
\newblock \href {https://doi.org/10.1007/s12532-023-00239-3} {\path{doi:10.1007/s12532-023-00239-3}}.

\bibitem{gurobioptimizationllcGurobiOptimizer2023}
{Gurobi Optimization, LLC}, {Gurobi Optimizer Reference Manual} (2023).

\bibitem{huangfuParallelizingDualRevised2018}
Q.~Huangfu, J.~A.~J. Hall, Parallelizing the dual revised simplex method, Mathematical Programming Computation 10~(1) (2018) 119--142.
\newblock \href {https://doi.org/10.1007/s12532-017-0130-5} {\path{doi:10.1007/s12532-017-0130-5}}.

\bibitem{malyutaConvexOptimizationTrajectory2022}
D.~Malyuta, T.~P. Reynolds, M.~Szmuk, T.~Lew, R.~Bonalli, M.~Pavone, B.~A{\c c}{\i}kme{\c s}e, Convex {{Optimization}} for {{Trajectory Generation}}: {{A Tutorial}} on {{Generating Dynamically Feasible Trajectories Reliably}} and {{Efficiently}}, IEEE Control Systems 42~(5) (2022) 40--113.
\newblock \href {https://doi.org/10.1109/MCS.2022.3187542} {\path{doi:10.1109/MCS.2022.3187542}}.

\bibitem{acikmeseConvexProgrammingApproach2007}
B.~Acikmese, S.~R. Ploen, Convex {{Programming Approach}} to {{Powered Descent Guidance}} for {{Mars Landing}}, Journal of Guidance, Control, and Dynamics 30~(5) (2007) 1353--1366.
\newblock \href {https://doi.org/10.2514/1.27553} {\path{doi:10.2514/1.27553}}.

\bibitem{vernerExplicitRungeKutta1978}
J.~H. Verner, Explicit {{Runge}}--{{Kutta Methods}} with {{Estimates}} of the {{Local Truncation Error}}, SIAM Journal on Numerical Analysis 15~(4) (1978) 772--790.
\newblock \href {https://doi.org/10.1137/0715051} {\path{doi:10.1137/0715051}}.

\bibitem{rackauckasDifferentialEquationsJlPerformant2017}
C.~Rackauckas, Q.~Nie, {{DifferentialEquations}}.jl -- {{A Performant}} and {{Feature-Rich Ecosystem}} for {{Solving Differential Equations}} in {{Julia}}, Journal of Open Research Software 5~(1) (2017) 15.
\newblock \href {https://doi.org/10.5334/jors.151} {\path{doi:10.5334/jors.151}}.

\bibitem{revelsForwardModeAutomaticDifferentiation2016}
J.~Revels, M.~Lubin, T.~Papamarkou, Forward-{{Mode Automatic Differentiation}} in {{Julia}} (Apr. 2016).

\bibitem{bernardiniStatedependentTrustRegion2024}
N.~Bernardini, N.~Baresi, R.~Armellin, State-dependent trust region for successive convex programming for autonomous spacecraft, Astrodynamics (Apr. 2024).
\newblock \href {https://doi.org/10.1007/s42064-024-0200-1} {\path{doi:10.1007/s42064-024-0200-1}}.

\end{thebibliography}
